%% file: joint.tex
\documentclass{article}

\usepackage[greek, english]{babel}
\usepackage[utf8]{inputenc}

\makeatletter
\renewcommand\paragraph{\@startsection{paragraph}{4}{\z@}
                                     {.75\baselineskip \@plus 1ex \@minus 1ex}
                                     {-0.5em  \@plus .2ex}
                                     {\normalsize\bfseries}}
\makeatother

\let\originalparagraph\paragraph
\renewcommand{\paragraph}[2][.]{\originalparagraph{#2#1}}

\usepackage{mathtools} 
\mathtoolsset{showonlyrefs=true}
\usepackage[ntheorem, framemethod=TikZ]{mdframed} 
\usepackage{framed}
\usepackage[amsmath, hyperref, framed]{ntheorem}%
\usepackage{amssymb}

\usepackage{bm}
\newcommand{\bmath}{\boldsymbol}
\input{mathematics} 

\newcommand\myatop[2]{\genfrac{}{}{0pt}{}{#1\hfill}{#2\hfill}}

\usepackage{graphicx}

\usepackage[labelsep=period, labelfont=bf]{caption}
\usepackage[labelformat=simple,labelsep=period,labelfont=bf]{subcaption}

\newcommand*{\myalign}[2]{\multicolumn{1}{#1}{#2}}

\usepackage{dcolumn}
\newcolumntype{d}[1]{D{.}{.}{#1}}

\usepackage[colorinlistoftodos,textwidth=1.7cm, textsize=footnotesize]{todonotes}

    \usepackage{lineno}

\newcommand*\patchAmsMathEnvironmentForLineno[1]{%
  \expandafter\let\csname old#1\expandafter\endcsname\csname #1\endcsname
  \expandafter\let\csname oldend#1\expandafter\endcsname\csname end#1\endcsname
  \renewenvironment{#1}%
     {\linenomath\csname old#1\endcsname}%
     {\csname oldend#1\endcsname\endlinenomath}}%
\newcommand*\patchBothAmsMathEnvironmentsForLineno[1]{%
  \patchAmsMathEnvironmentForLineno{#1}%
  \patchAmsMathEnvironmentForLineno{#1*}}%
\AtBeginDocument{%
\patchBothAmsMathEnvironmentsForLineno{equation}%
\patchBothAmsMathEnvironmentsForLineno{align}%
\patchBothAmsMathEnvironmentsForLineno{flalign}%
\patchBothAmsMathEnvironmentsForLineno{alignat}%
\patchBothAmsMathEnvironmentsForLineno{gather}%
\patchBothAmsMathEnvironmentsForLineno{multline}%
}

\usepackage{xr}
\usepackage{hyperref}%
\hypersetup{%
  breaklinks=true,%
  colorlinks=false,%
  linkcolor=blue,%
  citecolor=blue,%
  filecolor=blue,%
  urlcolor=blue%
}

\usepackage{ifthen}%
\usepackage{xkeyval}%
\usepackage{xfor}%
\usepackage{amsgen}%
\usepackage{etoolbox}%
\usepackage{datatool-base}%

\usepackage[%
  xindy,%
  style=long,%
  toc=true,%
  nonumberlist,%
  acronym,%
  acronymlists={abbreviations},%
  hyperfirst=false
  ]{glossaries}%

\input{abbreviations}

\theoremstyle{change}%

\qedsymbol{\ensuremath{\Box}}
\theoremstyle{change}%

\theoremheaderfont{\normalfont\bfseries\rmfamily} 
\theorembodyfont{\normalfont\slshape}%
\theoremseparator{.}

\theoremheaderfont{\normalfont\bfseries\rmfamily} 
\theorembodyfont{\normalfont\slshape}%

\theorembodyfont{\normalfont}%
\newmdtheoremenv[
 ntheorem=true,
 skipbelow = .6\baselineskip plus 1ex minus 1ex,
 skipabove = .6\baselineskip plus 1ex minus 1ex,
 innerleftmargin = 0pt,
 innerrightmargin = 0pt,
 leftline = false,
 rightline = false,
 needspace = 5ex 
]{framedAlgorithm}[theorem]{Algorithm}

\usepackage{csquotes}%
\usepackage{natbib}%
\setcitestyle{authoryear}%
\title{Efficient sequential Monte Carlo algorithms for integrated population models}

\author{Axel Finke$^{1}$, 
Ruth King$^{2,3}$, 
Alexandros Beskos$^{1}$, 
Petros Dellaportas$^{1,3,4}$\\[2ex] 
$^{1}$Department of Statistical Science, University College London, U.K. \\
$^{2}$School of Mathematics, University of Edinburgh, U.K.\\
$^{3}$The Alan Turing Institute, U.K.\\
$^{4}$Department of Statistics, Athens University of Economics and Business, Greece}

\usepackage[left=3.8cm, right=3.8cm, top=3cm, bottom=4cm]{geometry}

\begin{document}
 
\maketitle

\begin{abstract}
 State-space models are commonly used to describe different forms of ecological data. We consider the case of count data with observation errors. For such data the system process is typically multi-dimensional consisting of coupled Markov processes, where each component corresponds to a different characterisation of the population, such as age group, gender or breeding status. The associated system process equations describe the biological mechanisms under which the system evolves over time. However, there is often limited information in the count data alone to sensibly estimate demographic parameters of interest, so these are often combined with additional ecological observations leading to an integrated data analysis. Unfortunately, fitting these models to the data can be challenging, especially if the state-space model for the count data is non-linear or non-Gaussian. We propose an efficient \glsdesc{PMCMC} algorithm to estimate the demographic parameters without the need for resorting to linear or Gaussian approximations. In particular, we exploit the integrated model structure to enhance the efficiency of the algorithm. We then incorporate the algorithm into a \glsdesc{SMC} sampler in order to perform model comparison with regards to the dependence structure of the demographic parameters. Finally, we demonstrate the applicability and computational efficiency of our algorithms on two real datasets.\\[2ex]
\end{abstract}

\section{Introduction}
\glsunset{BUGS}
\glsunset{JAGS}

\emph{State-space models} are becoming an increasingly common and useful representation of many ecological systems \citep{buckland07, king14, newman14}. For example, they are often used to describe population count data \citep{newman98,besbeas2002integrating, king08}; telemetry data \citep{morales04,mcclintock12,breed12};  longitudinal growth data \citep{peters2010ecological}; fisheries data \citep{millar00}; capture-recapture and associated data \citep{dupuis95,royle08,king12}.

\paragraph{Inference in state-space models}
Unfortunately, fitting state-space models often leads to computational challenges because the likelihood -- expressible only as an integral or sum over the latent (\IE unobserved) states -- is typically intractable unless the states take values in some small, finite set or unless the model is linear and Gaussian in which case the likelihood can be evaluated via the Kalman filter \citep{kalman1960new,newman98}. Two approaches are typically applied to circumvent this problem.
\begin{itemize}
 \item The first is to approximate the state-space model with a model that is linear and Gaussian, \EG as in \citet{besbeas2002integrating}. 
 Unfortunately, such approximations introduce a bias which is often difficult to quantify. 
 \item The second is to impute the unobserved states alongside the model parameters within a \gls{MCMC} approach. 
 Unfortunately, such data-augmentation schemes -- in particular as normally implemented in \gls{BUGS} \citep{gilks1994language} or \gls{JAGS} \citep{plummer2003jags} (see \EG \citet{brooks04}) -- can be slow and poorly mixing if the system states and parameters are highly correlated because only (small) subsets of them are updated individually \citep{king11}.
\end{itemize}
To avoid the problems with these approaches, \citep{andrieu2009pseudo, andrieu2010particle} proposed \gls{PMCMC} algorithms (see \citet{knape12} for a recent application in ecology). These algorithms replace the intractable likelihood in the \gls{MH} algorithm with an unbiased estimate obtained through a \gls{SMC} algorithm (or ``\glsdesc{PF}''). \gls{PMCMC} methods do not require a modification of the model and, despite replacing the likelihood with an approximation, do not introduce bias. 

\paragraph{Integrated population models} 
In this work, count data are available on some species of interest, \IE estimates of population sizes over a set of discrete times \citep{besbeas2002integrating,besbeas2009completing,king08}. These count data are modelled as a state-space model to account for the measurement error. In addition to the count data, assume that other types of data are available on the species, \EG  capture-recapture, ring-recovery or nest-record data. To utilise all the available information to estimate demographic parameters of interest, it is necessary to combine these different data sets within a single \emph{integrated population model}. Unfortunately, actually fitting such models is challenging, in particular because they inherit all the above-mentioned difficulties with fitting the constituent state-space model.

\paragraph{Contributions}
In this work, we devise efficient methodology for performing full Bayesian parameter estimation and model comparison in integrated population models without the need for linear or Gaussian approximations to the state-space model. 
\begin{itemize}
  \item In Section~\ref{sec:parameter_estimation}, we first review standard \gls{PMCMC} methods for Bayesian parameter estimation in models with intractable likelihoods. Then, in Subsection~\ref{subsec:improving_mcmc_efficiency_for_integrated_models}, we exploit the structure of integrated population models to reduce the computational burden of the \gls{PMCMC} algorithm through a \glsdesc{DA} \citep{christen2005markov} technique.
  
  \item In Section~\ref{sec:model_comparison}, we first incorporate our \gls{PMCMC} methodology into 
  \gls{SMC} samplers \citep{chopin2002sequential, delmoral2006sequential, duan2015density, zhou2016towards} in order to estimate posterior model probabilities (or equivalently: Bayes factors) across a set of different integrated population models. This permits Bayesian model comparison without the need for reversible-jump \gls{MCMC} algorithms \citep{green1995reversible} which often mix poorly and can be difficult to implement and tune. Then, in Subsection~\ref{subsec:improving_smc_efficiency_for_integrated_models}, we again exploit the structure of integrated population models to reduce the computational burden of the \gls{SMC} sampler through a novel scheme which separately tempers the different likelihood terms.
  
  \item In Sections~\ref{sec:owls} and \ref{sec:herons}, we apply the proposed methodology to two real integrated data examples relating to (1) little owls and (2) grey herons. In both applications, our methodology yields reliable estimates of the model evidence, even in moderately high dimensions. 
  In the case of owls, we find that models proposed in the literature may be unnecessarily over-parametrised. For instance, we find no evidence for the hypothesis from \citet{abadi2010estimation} that the immigration rate of the owls depends on the abundance of voles -- their main prey.  We also demonstrate the utility of the \glsdesc{DA} approach. In the case of the herons, we show that the state-of-the-art models used in the literature, such as the threshold model from \citet{besbeas2012threshold}, fit poorly; to remedy this situation, we propose a novel regime-switching state-space model which significantly outperforms all existing models in terms of model fit and model evidence. 
\end{itemize}

\section{Integrated model}
\label{sec:integrated_modelling} 

\subsection{Data}

In this section, we combine multiple (independent) data sets, one of which being count data, obtained from a single population, within a single integrated model. Let $\dataCount = \{y_1,\dotsc,y_{\nObservationsCount}\}$ denote the \emph{count data} collected at times $t=1,\dotsc,\nObservationsCount$. Here, $y_t$ is the observation (subject to measurement error) of the true population size at time $t$. The observed counts may be multivariate, \EG counts for males and females; or juveniles and adults. Though in all the examples we consider later, the count data are univariate. Let $\dataOther$ denote all \emph{additional data} available such as capture-recapture data, ring-recovery data or nest-record data. The aim of this work is then to perform inference based on \emph{all data} $\data = \{\dataCount, \dataOther\}$.

\subsection{Likelihood}
\label{subsec:likelihood}

Let $\param \in \paramSet$ denote the collection of unknown model parameters. We assume that the collections of count and additional data are conditionally independent of each other given $\param$ so that the joint likelihood of the data can be written as a product of the individual likelihoods:
\begin{align}
  p(\data | \param)
  & = p(\dataCount | \param) p(\dataOther | \param). \label{eq:likelihood_factorisation} 
\end{align}
In Sections~\ref{sec:parameter_estimation} and \ref{sec:model_comparison}, we exploit this factorisation to enhance the efficiency of our proposed methodology. Throughout this work, we assume that the additional data $\dataOther$ are modelled in such way that $p(\dataOther | \param)$ can be evaluated pointwise.

A state-space model is specified for the count data. Let $\latentFull = \{\latentVec_1, \dotsc, \latentVec_{\nObservationsCount}\} \in \latentSet^T$ (for some relevant space $\latentSet$) denote the true (unobserved) population counts with initial density $\mu_\param(\latentVec_1)$ and transitions $f_\param(\latentVec_t|\latentVec_{t-1})$. Furthermore, let $g_\param(y_t|\latentVec_t)$ be the density of the $t$th observed count given $\latentVec_t$. Then, conditional on $\param$, the joint distribution of $\dataCount$ and $\latentFull$ is: 
\begin{align}
  p(\dataCount, \latentFull |\param) & = \mu_\param(\latentVec_1) g_\param(y_1|\latentVec_1) \prod_{t=2}^{T} f_\param(\latentVec_t|\latentVec_{t-1}) g_\param(y_t|\latentVec_t). \label{eq:pf_target_unnormalised}
\end{align}
The (marginal) count-data likelihood is thus given by the integral (or sum, if $\latentSet$ is discrete)
\begin{align}
  p(\dataCount|\param) = \int_{\latentSet^T} p(\dataCount, \latentFull |\param) \intDiff \latentFull. \label{eq:marginal_likelihood_count_data}
\end{align}
Throughout this work, we assume that this integral (sum) is intractable as is usually the case unless $\latentSet$ is finite and sufficiently small or unless the state-space model is linear and Gaussian in which case \eqref{eq:marginal_likelihood_count_data} can be evaluated using the Kalman filter.

\subsection{Posterior distribution}
\label{subsec:posterior}

Let $p(\param)$ denote the prior distribution of the parameters then the (marginal) \emph{posterior} distribution of the parameters $\param$ (given the full data $\data$) is given by
\begin{align}
  \pi(\param) 
  & \coloneqq p(\param|\data) = \frac{p(\data | \param) \paramPrior(\param)}{p(\data)},
  \label{eq:posterior}
\end{align}
where
\begin{align}
 p(\data) \coloneqq \int_\paramSet p(\data|\param) p(\param) \diff \param,\label{eq:evidence}
\end{align}
in the denominator is the \emph{evidence} for the model. This quantity plays a key r\^ole in Bayesian model comparison as outlined in Section~\ref{sec:model_comparison}. The posterior distribution is typically intractable as the integrals in \eqref{eq:marginal_likelihood_count_data} and \eqref{eq:evidence} are not of closed form. Instead, we approximate this distribution via Monte Carlo methods as described in the next section.

\section{Parameter estimation}
\label{sec:parameter_estimation}

\glsreset{MCMC}
\glsreset{MH}
\glsreset{SMC}
\glsreset{PF}
\glsreset{PMCMC}

\subsection{Particle MCMC}
\label{subsec:pmcmc}

In this section, we describe \gls{MCMC} methods for approximating the posterior distribution of the model parameters. We also propose modifications which exploit the structure of integrated models to improve efficiency of the algorithm. For now, we assume that the model is known -- model uncertainty is dealt with in Section~\ref{sec:model_comparison}.

As the count-data likelihood $p(\dataCount|\param)$ (and hence the overall likelihood $p(\data|\param)$) is intractable, we cannot implement the \emph{idealised} \gls{MH} algorithm which, at each iteration,  proposes new set of parameters $\varParam \sim q(\varParam|\param)$ and then accepts it \gls{WP} $\min\{1,r\}$, where
\begin{align}
 r \coloneqq \frac{q(\param|\varParam)}{q(\varParam|\param)} \frac{\paramPrior(\varParam)}{\paramPrior(\param)} \frac{p(\data|\varParam)}{p(\data|\param)}. 
\end{align}
A common solution is to use a data-augmentation approach which imputes the latent variables $\latentFull$ (alongside the parameters) within the algorithm. However, the number states is typically large so that single-site updates are then required. This approach, commonly used in ``black-box'' samplers such as \gls{BUGS} or \gls{JAGS}, can lead to poor mixing if highly correlated variables or parameters are updated separately. 

To avoid such problems, we employ \gls{PMCMC} algorithms \citep{andrieu2010particle}. These replace $p(\dataCount|\param)$ in the acceptance ratio of the idealised \gls{MH} algorithm with an unbiased estimate $\hat{p}(\dataCount|\param)$ obtained through \gls{SMC} methods. Crucially, the resulting algorithm still targets the correct posterior distribution. 

Before stating the \gls{PMCMC} algorithm, we review \gls{SMC} algorithms. A comprehensive discussion of the application  of \gls{SMC} algorithms to state-space models -- usually termed \glspl{PF} in this setting -- can be found in \citet{cappe2005inference, doucet2011tutorial}. A simple \gls{PF} is outlined in Algorithm~\ref{alg:pf}, where we use the convention that actions prescribed for the $n$th particle are to be performed conditionally independently for all $n \in \{1, \dotsc, \nParticlesLower\}$.

\begin{flushleft}
  \begin{framedAlgorithm}[particle filter] \label{alg:pf}~
    \begin{flushleft}
      \begin{enumerate}
       	\item At Step~$1$, sample $\smash{\latentVec_1^n \sim \mu_{\param}}(\latentVec_1)$ and set $\smash{\weightLower_1^n \coloneqq g_\param(y_1|\latentVec_1^n)}$,
	\item\label{alg:smc:stept} At Steps~$t = 2,\dotsc,\nObservationsCount$, 
	\begin{enumerate}
	  \item\label{alg:smc:resampling} sample $\smash{\ancestorIndexLower_{t-1}^{n} = l}$ \WP $\smash{\WeightLower_{t-1}^l \coloneqq \weightLower_{t-1}^l / \sum_{k=1}^\nParticlesLower \weightLower_{t-1}^k}$, 
	  \item sample $\latentVec_t^n \sim f_{\param}(\latentVec_t|\latentVec_{t-1}^{a_{t-1}^n})$ and set $\smash{\weightLower_t^n \coloneqq g_\param(y_t|\latentVec_t^n)}$. 
	\end{enumerate}
      \end{enumerate}
    \end{flushleft}
  \end{framedAlgorithm}
\end{flushleft}
At the end of Algorithm~\ref{alg:pf}, an unbiased \citep{delmoral1996nonlinear} estimate of $p(\dataCount|\param)$ is given by
\begin{align}
 \hat{p}(\dataCount|\param) \coloneqq \prod_{t=1}^\nObservationsCount \frac{1}{\nParticlesLower} \sum_{n=1}^\nParticlesLower \weightLower_t^n. \label{eq:pf_marginal_likelihood_estimate}
\end{align}
Numerous extensions exist for making Algorithm~\ref{alg:pf} more efficient. The particular version of \gls{PF} we use in our applications is outlined in Web Appendix~C. 

We now describe the \gls{PMCMC} algorithm. A single \gls{PMCMC} update is outlined in Algorithm~\ref{alg:pmmh}, where $\invTemp \in [0,1]$ is a parameter which will be used by the evidence-approximation algorithms in Section~\ref{sec:model_comparison}. For the moment, simply take $\invTemp = 1$.

\begin{flushleft}
  \begin{framedAlgorithm}[particle MCMC] \label{alg:pmmh} 
    \begin{flushleft}
      At each iteration, given $(\param, \hat{p}(\dataCount|\param))$,
      \begin{enumerate}
	\item propose $\varParam \sim q(\varParam|\param)$,
	\item generate $\hat{p}(\dataCount|\varParam)$ using Alg.~\ref{alg:pf} (with $\param = \varParam$),
	\item return $(\varParam, \hat{p}(\dataCount|\varParam))$ \gls{WP} $\min\{1,r\}$, where
	\begin{align}
	 r \coloneqq \frac{q(\param|\varParam)}{q(\varParam|\param)} \frac{\paramPrior(\varParam)}{\paramPrior(\param)} \biggl[\frac{\hat{p}(\dataCount|\varParam)p(\dataOther|\varParam)}{\hat{p}(\dataCount|\param)p(\dataOther|\param)}\biggr]^{\mathrlap{\alpha}};
	\end{align}
        otherwise, return $(\param, \hat{p}(\dataCount|\param))$.
      \end{enumerate}
    \end{flushleft}
  \end{framedAlgorithm}
\end{flushleft}

\subsection{Improving PMCMC efficiency for integrated models}
\label{subsec:improving_mcmc_efficiency_for_integrated_models}

The computational cost of the \gls{PMCMC} update in Algorithm~\ref{alg:pmmh} is dominated by the \gls{PF} used to evaluate the estimate of $p(\dataCount|\varParam)$ for each proposed parameter value $\varParam$. To improve the efficiency of algorithm, we utilise the  propose a \gls{DA} approach \citep{christen2005markov,sherlock2015efficiencyHapalike} based on the factorisation of the likelihood function in \eqref{eq:likelihood_factorisation}. The idea is to avoid invoking the \gls{PF} for proposed values $\varParam$ which are not compatible with the additional data $\dataOther$ and which are therefore likely to be rejected in Algorithm~\ref{alg:pmmh}. This can improve efficiency if $\dataOther$ is highly informative about a large proportion of the model parameters. \gls{DA} was previously combined with \gls{PMCMC} updates in \citet{golightly2015delayed} (though in a slightly different way). Algorithm~\ref{alg:pmmh_delayed_acceptance} summarises the approach whose validity may be established using the arguments of \citet{christen2005markov, andrieu2010particle}. Again, assume for the moment that $\alpha=1$.

\begin{flushleft}
  \begin{framedAlgorithm}[delayed acceptance PMCMC] \label{alg:pmmh_delayed_acceptance} 
    \begin{flushleft}
      At each iteration, given $(\param, \hat{p}(\dataCount|\param))$,
      \begin{enumerate}
	\item propose $\varParam \sim q(\varParam|\param)$,
	\item go to Step~\ref{alg:pmmh_delayed_acceptance:3} \gls{WP} $\min\{1,r\}$, where 
	\begin{align}
	  r \coloneqq \frac{q(\param|\varParam)}{q(\varParam|\param)} \frac{\paramPrior(\varParam)}{\paramPrior(\param)} \biggl[\frac{p(\dataOther|\varParam)}{p(\dataOther|\param)}\biggr]^{\mathrlap{\alpha}};
	\end{align}
	otherwise, return $(\param, \hat{p}(\dataCount|\param))$.
	\item \label{alg:pmmh_delayed_acceptance:3} Generate $\hat{p}(\dataCount|\varParam)$ using Alg.~\ref{alg:pf} (with $\param = \varParam$),
	\item return $(\varParam, \hat{p}(\dataCount|\varParam))$ \gls{WP} $\min\{1,r\}$, where 
	\begin{align}
	 r \coloneqq \biggl[\frac{\hat{p}(\dataCount|\varParam)}{\hat{p}(\dataCount|\param)}\biggr]^{\mathrlap{\alpha}};
	\end{align}
        otherwise, return $(\param, \hat{p}(\dataCount|\param))$.
      \end{enumerate}
    \end{flushleft}
  \end{framedAlgorithm}
\end{flushleft}


\section{Model comparison}
\label{sec:model_comparison}


\subsection{Posterior model probabilities}

Let $\{\modelIndicator_i \colon i \in \modelSet\}$ denote some finite collection of plausible biological models of interest. To indicate the $i$th model, we now add the model indicator $\modelIndicator_i$ to the densities from Section~\ref{sec:integrated_modelling}. That is, the prior of the parameters $\param \in \paramSet_i$ is now written as $p(\param|\modelIndicator_i)$, the likelihood as $\smash{p(\data|\param, \modelIndicator_i) = p(\dataCount|\param, \modelIndicator_i)p(\dataOther|\param, \modelIndicator_i)}$ and the evidence as $p(\data |\modelIndicator_i) = \int_{\paramSet_i} p(\data|\param, \modelIndicator_i) p(\param|\modelIndicator_i) \intDiff \param$.

Let $p(\modelIndicator_i)$ denote the prior probability of the $i$th model. Bayesian model comparison is based on the \emph{posterior model probabilities} \citep[Chapter~6]{bernardo2001bayesian}
\begin{align}
  p(\modelIndicator_i|\data) \coloneqq \frac{p(\modelIndicator_i)  p(\data|\modelIndicator_i)}{\sum_{j \in \modelSet} p(\modelIndicator_j) p(\data|\modelIndicator_j)}. \label{eq:model_probability}
\end{align}
Unfortunately, the model evidence $p(\data |\modelIndicator_i)$ and hence the posterior model probabilities in \eqref{eq:model_probability} are intractable. To perform model comparison, we replace the model evidence $p(\data | \modelIndicator_i)$ with an estimate $\hat{p}(\data | \modelIndicator_i)$ obtained via an \gls{SMC} sampler. As a by-product, the \gls{SMC} sampler also yields an approximation of the posterior distribution of $\param$ under the $i$th model. 



\subsection{SMC sampler for evidence approximation}
\label{subsec:smc_sampler}

For the moment, assume that $p(\dataCount|\param, \modelIndicator_i)$ can be evaluated. A simple importance-sampling approximation of $\smash{p(\data|\modelIndicator_i)}$ is then given by $\smash{\frac{1}{\nParticlesUpper}\sum_{m=1}^\nParticlesUpper p(\data|\param^m,\modelIndicator_i)}$, where $\param^1, \dotsc, \param^\nParticlesUpper$ are sampled independently from $p(\param|\modelIndicator_i)$. However, this approach typically performs poorly if there is a strong mismatch between the prior and the posterior (which is common, especially if $\param$ is high-dimensional or if the data are highly informative). 
To circumvent this problem, we employ an \gls{SMC} sampler \citep{chopin2002sequential, delmoral2006sequential} which uses successive importance-sampling steps to approximate a \emph{sequence} of distributions to smoothly bridge the gap between the prior and the posterior,
\begin{align}
 \paramPrior(\param|\modelIndicator_i) = \extendedTarget_0(\param),\extendedTarget_1(\param),\dotsc, \extendedTarget_\nStepsUpper(\param) = p(\param|\data, \modelIndicator_i). \label{eq:smc_sampler_sequence_of_target_distributions}
\end{align}
The idea behind \gls{SMC} samplers is that each individual importance-sampling step (\IE proposing samples from $\extendedTarget_{s-1}(\param)$ to approximate $\extendedTarget_s(\param)$) may be feasible even if the gap between the prior $\extendedTarget_0(\param)$ and the posterior $\extendedTarget_\nStepsUpper(\param)$ is wide. We use a likelihood-tempering approach,
\begin{align}
 \extendedTarget_s(\param) \propto \paramPrior(\param|\modelIndicator_i) p(\data|\param,\modelIndicator_i)^{\invTemp_s}, \label{eq:smc_sampler_target_distribution_at_step_s}
\end{align}
where the \emph{temperatures} $0 = \invTemp_0 < \invTemp_1 < \dotsc < \invTemp_\nStepsUpper = 1$ (and the number of bridging distributions, $\nStepsUpper$) can then be tuned to ensure that the interpolation between the prior and posterior in \eqref{eq:smc_sampler_sequence_of_target_distributions} is sufficiently smooth. Of course, in the models considered in this work, $p(\dataCount|\param)$ is intractable and is therefore again approximated using a \gls{PF}. This idea was first employed by \citet{duan2015density} and it shares some similarities with the \gls{SMC}\textsuperscript{2} approach from \citet{chopin2013smc2} which we discuss at the end of this section.

Algorithm~\ref{alg:smc_sampler_intractable_marginal_likelihood} outlines the \gls{SMC} sampler; we use the convention that any action specified for the $m$th particle is to be performed conditionally independently for \emph{all} $m \in \{1,\dotsc,\nParticlesUpper\}$.
\begin{flushleft}
\begin{framedAlgorithm}[SMC sampler]~ \label{alg:smc_sampler_intractable_marginal_likelihood}
\begin{enumerate}
 \item At Step~$0$, 
 \begin{enumerate}
   \item sample $\smash{\param_0^m  \sim \paramPrior(\param|\modelIndicator_i)}$, 
   \item generate $\hat{p}_0^m(\dataCount|\param_0^m,\modelIndicator_i)$ using Alg.~\ref{alg:pf} (with $\param = \param_0^m$).
 \end{enumerate}
 \item At Step~$s = 1, \dotsc, \nStepsUpper$,
 \begin{enumerate}
  
  \item write $u_{s-1}^m \coloneqq \hat{p}_{s-1}^m(\dataCount|\param_{s-1}^m, \modelIndicator_i)p(\dataOther|\param_{s-1}^m, \modelIndicator_i)$,
  \item set $\smash{\weightUpper_s^m \coloneqq (u_{s-1}^m)^{\invTemp_s-\invTemp_{s-1}}}$, 

  \item sample $\smash{\ancestorIndexUpper_{s-1}^{m} = l}$ \gls{WP} $\smash{\WeightUpper_s^l \coloneqq \weightUpper_s^l / \sum_{k=1}^\nParticlesUpper \weightUpper_s^k}$,
  \item \label{alg:smc_sampler_intractable_marginal_likelihood:pmmh} sample $(\param_s^m, \hat{p}_s(\dataCount|\param_{s}^m, \modelIndicator_i))$ using Alg.~\ref{alg:pmmh}\\ (with $\alpha = \alpha_{s}$; $\param=\param_{s-1}^{\ancestorIndexUpper_{s-1}^m}$; $\smash{\hat{p}(\dataCount|\param)=\hat{p}_{s-1}^{\ancestorIndexUpper_{s-1}^m}(\dataCount|\param_{s-1}^{\ancestorIndexUpper_{s-1}^m},\modelIndicator_i)}$).
 \end{enumerate}
\end{enumerate}
\end{framedAlgorithm}
\end{flushleft} 
We then approximate the evidence
$p(\data|\modelIndicator_i)$ by
\begin{align}
\hat{p}(\data|\modelIndicator_i) \coloneqq \prod_{s=1}^\nStepsUpper \frac{1}{\nParticlesUpper}\sum_{m=1}^\nParticlesUpper \weightUpper_s^m. \label{eq:approximating_model_evidence}
\end{align}
The output of the algorithm can also be used to infer the model parameters in the $i$th model. That is, any posterior expectation $\smash{\E[\testFun(\param)]}$, for $\param \sim p(\param|\data,\modelIndicator_i)$ and where $\testFun$ is some test function, can be approximated by $\sum_{m=1}^\nParticlesUpper \WeightUpper_\nStepsUpper^m \testFun(\param_\nStepsUpper^m)$. Numerous extensions exist for making Algorithm~\ref{alg:smc_sampler_intractable_marginal_likelihood} more efficient. The particular version of \gls{SMC} sampler we use in our applications is outlined in Web Appendix~C. 
 
Other methods for performing model comparison using \gls{SMC} samplers can be found in \citet{zhou2016towards} (see also \citet{jasra2008interacting} for some extensions). In addition, \citet{chopin2013smc2} proposed another special case of the \gls{SMC}-sampler framework from \citet{delmoral2006sequential}, called \emph{\gls{SMC}\textsuperscript{2}}. This algorithm can be useful when one wishes to perform inference \emph{sequentially} because it can incorporate new data points as they arrive. However, as observed in \citet{drovandi2016alive}, the \gls{SMC}\textsuperscript{2} algorithm can become unstable when the newly-arrived observation contains information about the parameters which contradicts the existing information. In such cases, the likelihood-tempering approach adopted here can lead to a smoother sequence of target distributions \citet{duan2015density} and hence more accurate estimates.
In addition, \gls{SMC}\textsuperscript{2} does not easily accommodate variance-reductions schemes from, for example, \citet{gramacy2010importance} and \citet{nguyen2015efficient}.

\subsection{Improving SMC efficiency for integrated models}
\label{subsec:improving_smc_efficiency_for_integrated_models}

We are able to exploit the structure of integrated population models to enhance the efficiency of the \gls{SMC} sampler for evidence approximation. Firstly, we employ the \gls{DA} approach from Subsection~\ref{subsec:improving_mcmc_efficiency_for_integrated_models} to reduce the computational cost of the \gls{MCMC} updates in the \gls{SMC} sampler.
Secondly, we propose to employ a likelihood-tempering approach which tempers the different parts of the likelihood separately. That is, for some $1 \leq \nStepsUpperFirst < \nStepsUpper$, the \gls{SMC} sampler targets the distributions 
\begin{align}
 \extendedTarget_s(\param) \propto 
 \begin{cases}
   \paramPrior(\param|\modelIndicator_i) p(\dataOther|\param,\modelIndicator_i)^{\invTempFirst_s}, & \text{if $0 \leq s \leq \nStepsUpperFirst$,}\!\!\!\!\!\\
   \paramPrior(\param|\modelIndicator_i) p(\dataOther|\param,\modelIndicator_i) p(\dataCount|\param,\modelIndicator_i)^{\invTempSecond_s}, & \text{if $\nStepsUpperFirst < s \leq \nStepsUpper$,}\!\!\!\!\!\\
 \end{cases}
 \label{eq:smc_sampler_double_tempering_target_distribution_at_step_s}
\end{align}
where $0 = \invTempFirst_0 < \invTempFirst_1 < \dotsc < \invTempFirst_{\nStepsUpperFirst} = 1$ and $0 < \invTempSecond_{\nStepsUpperFirst+1} < \dotsc < \invTempSecond_{\nStepsUpper} = 1$. Of course, the intractable count-data likelihood is again replaced by an unbiased estimate. The advantage of this refined tempering scheme is that the approximation of the count-data likelihood (obtained through the costly \gls{PF}) is not needed in the first $\nStepsUpperFirst$ steps of the algorithm so that $\nStepsUpperFirst$ can be taken to be large. Introducing the additional data likelihood first can be especially beneficial if the additional data are highly informative about the parameters relative to the count data. This refined tempering strategy was crucial for obtaining reliable estimates in the herons example from Section~\ref{sec:herons} and its efficiency gains are also illustrated in Web Appendix A.

\section{Example 1: Little owls}
\label{sec:owls}

\subsection{Data}

In this section, we consider little-owl data described by \citet{schaub2006local} and subsequently analysed in \citet{abadi2010estimation}. 
The count data represent the number of breeding females at nest boxes near G\"oppingen, Southern Germany, observed anually from 1978 to 2003 (\IE $\nObservationsCount = 26$). The nest boxes were checked multiple times annually and data were recorded relating to overall population size (number of occupied nest boxes and number of breeding females); capture-recapture histories of individuals observed at nest boxes and reproductive success of the nests. In addition, time-varying covariate information about the abundance of voles  -- the primary prey for little owls -- is available. For further details for the study see \citet{schaub2006local}. 


\subsection{Parameters}

The main model parameters -- potentially specific to age group $a \in \{\juvenile,\adult\}$ ($\juvenile$: juvenile, \IE first-year, $\adult$: adult) and gender $g \in \{\male, \female\}$ ($\female$: female, $\male$: male) of the owls, and to time index $t\in \{1,\dotsc, \nObservationsCount\}$ -- are
\begin{description}
  \item[$\phi_{a,g,t}$:] probability of an owlof gender $g$ surviving until time $t+1$ if alive and aged $a$ at time $t$;
  \item[$p_{g,t+1}$:] probability of observing an owl of gender $g$ at time $t+1$ if alive at time $t+1$;
  \item[$\rho_t$:] productivity rate governing the number of chicks produced per female at time $t$ that survive to fledgling;
  \item[$\eta_t$:] immigration rate governing the number of female immigrants at time $t+1$ per female of the population at time $t$.
\end{description}

\subsection{Model specification}

We consider the model defined by \citet{schaub2006local} and subsequently fitted in \gls{BUGS} by \citet{abadi2010estimation} -- for further information and biological rationale see these papers. 

\subsubsection{Count-data model}

The system process, in terms of the true population sizes for the juvenile and adult females, $\latentVec_t = \{\latent_{\juvenile,t}, \latent_{\adult,t}\}$, is described by
\begin{align}
  \latent_{\juvenile,t+1} | \latentVec_t, \param 
  &\sim \dPois\bigl([\latent_{\juvenile,t}+\latent_{\adult,t}] \rho_t \phi_{1,\female,t}/2\bigr),\\
  \latent_{\adult,t+1} 
  &= \survivors_{t+1} + \immigrants_{t+1},
\end{align}
where $\survivors_{t+1}|\latentVec_t,\param \sim \dBin(\latent_{\juvenile,t}+\latent_{\adult,t}, \phi_{\adult,\female,t})$ is the number of female adults which survive from time $t$ to time $t+1$, and $\immigrants_{t+1}|\latentVec_t,\param  \sim  \dPois\bigl([\latent_{\juvenile,t}+\latent_{\adult,t}] \eta_t)$ is the number of female adults which immigrate in this period. We take the initial population sizes $\latent_{\juvenile,1}$ and $\latent_{\adult,1}$ to be a-priori independently distributed according to a discrete uniform distribution on $\{0,1,\dotsc,50\}$. Finally, the observation process is specified by $y_t|\latentVec_t, \param \sim \dPois(\latent_{\juvenile,t}+\latent_{\adult,t})$.

\subsubsection{Capture-recapture model}


Capture-recapture data are available in the form of the age-group and gender specific matrices $\dataRecapture \coloneqq \{\dataRecapture_{a,g}\colon a \in \{\juvenile, \adult\}, g \in \{\male, \female\}\}$. The $t$th row, denoted $\dataRecapture_{a,g,t} \coloneqq \{m_{a,g,t,s} \colon 1 < s \leq \nObservationsCount+1\}$, corresponds to the $t$th year of release ($t \in \{1,\dotsc,\nObservationsCount-1\}$). That is, $m_{a,g,t,s}$ is the number of individuals of gender $g$, last observed at age $a$ at time $t$, that are recaptured at time $s$ (if $t+1 \leq s \leq \nObservationsCount$) or never recaptured again (if $s = \nObservationsCount+1$). Note that $m_{a,g,t,s} = 0$ if $s \leq t$. 
For each year of release, we assume a multinomial distribution for the subsequent recaptures. The capture-recapture model specified as
\begin{align}
  \dataRecapture_{a,g,t} | R_{a,g,t}, \param & \sim \dMult(R_{a,g,t},\mathbf{q}_{a,g,t}).
\end{align}
Here, $R_{a,g,t}$ denotes the number of owls in age group $a$ and of gender $g$ that are recorded as being observed (either an initial capture or, if $a = \adult$, as a recapture) at time~$t$ and subsequently released. The multinomial cell probabilities $\mathbf{q}_{a,g,t} \coloneqq \{q_{a,g,t,s} \colon 1 < s \leq \nObservationsCount+1\}$ are given by 
\begin{align}
  q_{a,g,t,s} \coloneqq 
  \begin{cases}
    0, & \text{if $1 < s \leq t$,}\\
    \phi_{a,g,t} p_{g,s} 
    \prod_{r=t+1}^{s-1} \phi_{\adult,g,r} (1-p_{g,r}), & \text{if $t < s  \leq \nObservationsCount$,}\\
    1-\sum_{r=1}^{\nObservationsCount} q_{a,g,t,r}, & \text{if $s = \nObservationsCount+1$.}
  \end{cases}
\end{align}

\subsubsection{Fecundity model}
Nest record data  $\mathbf{n} \coloneqq \{N_t, n_t \colon 1 \leq t \leq \nObservationsCount\}$ are also available to provide information relating to the fecundity rate of little owls. Specifically, $N_t$ is the number of chicks that are produced at time $t$ and $n_t$ is the number of these chicks that survive to leave the nest. Following \citet{schaub2006local} we specify $n_t|N_t, \param \sim \dPois(N_t \rho_t)$.
With this notation, the set of all additional data is $\dataOther = \{\dataRecapture, \dataFecundity\}$.

\subsection{Parametrisation and Priors}

There is additional covariate information about the abundance of voles -- the primary source of prey for little owls -- at the study site, classified as low ($\voleCovar_t = 0$) or high ($\voleCovar_t = 1$), for each year of the study. Following \citet{schaub2006local, abadi2010estimation}, we parametrise
\begin{align}
  \logit \phi_{a,g,t} & = \alpha_0 + \alpha_1 \ind\{g = \male\} + \alpha_2 \ind\{a = \adult\} + \alpha_3 \yearCovar_t,\\
  \log \eta_t         & = \delta_0 + \delta_1 \voleCovar_t,\\
  \logit p_{g,t+1}    & = \beta_1 \ind\{g = \male\} + \beta_{t+1},
\end{align}
for $t = 1, \dotsc, \nObservationsCount-1$, where the additional covariate $\yearCovar_t$ denotes the normalised year. Furthermore, we specify $\log \rho_t = \gamma_t$, for $1 \leq t\leq  \nObservationsCount$.

We assume that all the model parameters in $\param$ are independent a-priori with a $\dN(-2,2)$ prior on $\delta_0$; all remaining parameters have $\dN(0,2)$ priors.

\subsection{Results}
\glsreset{DA}

We illustrate the performance gains obtained through the \gls{DA} approach within the \gls{MCMC} algorithm for parameter estimation. For simplicity, we only report results for the case that the productivity rate is constant over time and with immigration independent of the abundance of voles, \IE $\gamma_1 = \dotsc = \gamma_\nObservationsCount$ and $\delta_1 = 0$, as this was one of the specifications which performed best in terms of model evidence. Other tested model specifications are listed in Web Appendix~A and the estimated model evidence for each is reported in Web Figure~1. This figure also illustrates that the data do not support the hypothesis, proposed in \citet{abadi2010estimation}, that little-owl immigration depends on the abundance of voles.

\begin{figure}[!ht]
 \centering
  \includegraphics[scale=1]{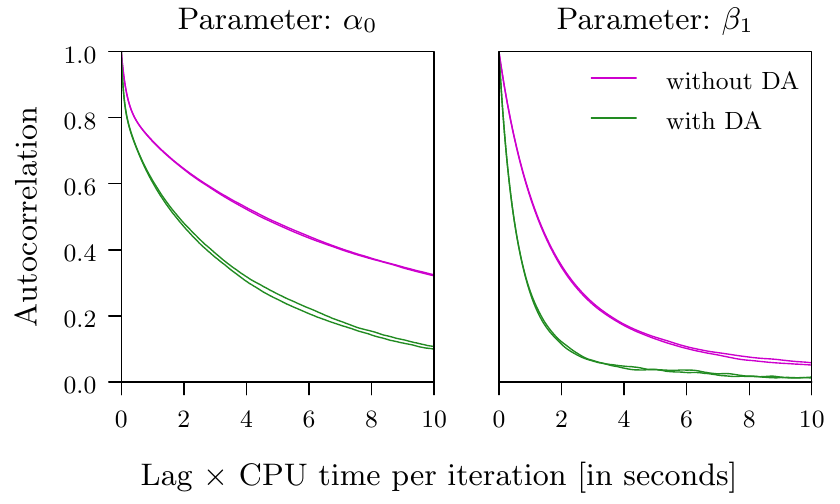}
\vspace{-0.2cm}
 \caption{Autocorrelation (rescaled by computation time) of the estimates of the parameters $\alpha_0$ and $\beta_1$ in the little-owls model (with the productivity rates assumed to be constant, \IE $\gamma_1 = \dotsc = \gamma_\nObservationsCount$) and immigration independent of the abundance of voles, \IE $\delta_1 = 0$. The results are based on two independent repeats (each comprised of $10^7$ iterations) of the \gls{MCMC} algorithms with and without \glsdesc{DA}.}
  \label{fig:owls_acf}
\end{figure}

Figure~\ref{fig:owls_acf} illustrates the utility of our proposed \gls{DA} approach. It shows that even though \gls{DA} decreases the acceptance rate, the computational savings attained by only invoking the \gls{PF} for ``promising'' parameters more than compensate for this. Further efficiency gains due to the refined tempering scheme from Subection~\ref{subsec:improving_smc_efficiency_for_integrated_models} are demonstrated in Appendix A.

\section{Example 2: Grey herons}
\label{sec:herons}

\subsection{Data}

In this section, we consider grey-heron data previously presented and analysed by \citet{besbeas2002integrating,besbeas2009completing,besbeas2012threshold}. The count data $\dataCount$ correspond to the estimated number of female herons (or breeding pairs) in the UK, from 1928 to 1998, \IE for $\nObservationsCount = 71$ time periods. Within our application we also have ring-recovery data for individuals released between 1955 and 1997.

\subsection{Parameters}

Following \citet{besbeas2009completing} we specify up to four age categories for the herons in order to account for different survival probabilities, with younger herons typically having a lower survival probability than older adults. We indicate the age group by the subscript $a \in  \{1,\dotsc,\adult\}$, where $a=1$ represents first-years, $a=2$ represents second-years, etc.\ while $a=\adult$ represents all the remaining adults. The main model parameters are then
\begin{description}
  \item[$\phi_{a,t}$:] probability of a heron surviving until time $t+1$ if alive and aged $a$ at time $t$;
  \item[$\rho_t$:] productivity rate governing the number of females produced per female at time $t$;
  
  \item[$\lambda_t$:] probability of recovering a dead heron in $[t,t+1)$ if it died in that interval.
\end{description}

\subsection{Model specification}

We follow \citet{besbeas2009completing} with regard to the model specification, allowing for some judicial changes in the state-space model specification. 

\subsubsection{Count-data model}

We once again specify state-space model for the count data $\dataCount = \{y_1,\dotsc,y_\nObservationsCount\}$. We let $\latent_{a,t}$, denote the true population sizes of herons in age group $a$ at time $t$. The system process is then  described by
\begin{align}
  \latent_{t,1}|\latentVec_{t-1},\param & \sim \smash{\textstyle\dPois( \rho_{t-1} \phi_{1,t-1}\sum_{a=2}^{\adult}\latent_{a,t-1})},\\*
  \latent_{a,t}|\latentVec_{t-1},\param & \sim \dBin(\latent_{a-1,t-1}, \phi_{a,t-1}), \quad \text{for $1 < a < \adult$,}\!\!\!\!\!\\*
  \latent_{\adult,t}|\latentVec_{t-1},\param & \sim \dBin(\latent_{\adult-1,t-1} + \latent_{\adult,t-1}, \phi_{\adult,t-1}).
\end{align}
For simplicity, we assume that the distribution of each component of the initial state is a negative-binomial distribution with probability $p=1/100$ and size $n_0 = \mu_0 p/(1-p)$ for age groups $1 \leq a < \adult$ and $n_1 = \mu_1 p/(1-p)$ for adults, respectively. We specify the means $\mu_0 = 5000/5$ and $\mu_1=5000 - (\adult-1) \mu_0$ in such a way that a-priori, $\E[\sum_{a=1}^A \latent_{a,1} |\param] = 5000$.


Such state-space model are typically approximated by a linear-Gaussian model in order to permit inference via the Kalman filter \citep{besbeas2002integrating}. However, the assumption that $g_\param(y_t|\latentVec_t)$ is Gaussian is typically unrealistic, since it implies that the observation error is independent of scale and continuous. Alternatively, to incorporate the effect of scale a lognormal distribution has been applied \citep{king08}, but this too assumes a continuous distribution for the discrete observations.  
Instead, we consider a negative-binomial observation process (with probability/size parametrisation), such that 
\begin{align}
  y_t | \latentVec_t,\param \sim \dNegBin\biggl(\frac{\kappa}{1-\kappa}\sum_{a=2}^A \latent_{a,t},\kappa\biggr), \label{eq:negative_binomial_observation_density}
\end{align}
for some $\kappa \in (0,1)$. This specification permits overdispersed observations since $\E[y_t|\latentVec_t,\param] = \sum_{a=2}^A \latent_{a,t} > \sum_{a=2}^A \latent_{a,t}/\kappa = \var[y_t|\latentVec_t,\param]$. 

\subsubsection{Ring-recovery data model}
Recall that count data are available from 1928 to 1998, \IE for $\nObservationsCount = 71$ time periods. In contrast, ring-recovery data are only available for individuals released between 1955 and 1997, \IE released in time period $t\in\{t_1,\dotsc,t_2\}$, where $t_1 = 28$ and $t_2 = 70$. These data are stored in a matrix $\dataOther$ whose $t$th row is denoted $\dataOther_t = \{w_{t,s}\colon t_1+1 \leq s \leq t_2+2\}$. Here, $w_{t,s}$ indicates the number of individuals released at time $t$ which are subsequently recovered dead in the interval $(s-1,s]$; $w_{t,t_2+2}$ corresponds to the number of individuals that are released at time $t$ that are not seen again within the study. 

For each year of release, we assume a multinomial distribution for the subsequent recoveries (see \EG \citet{mccrea14} for further explanations on the ring-recovery model). Thus, the model for $\dataRing$ is then specified as
\begin{align}
  \dataOther_t | R_t, \param \sim \dMult(R_t, \mathbf{q}_t).
\end{align}
Here, $R_t$ denotes the number of herons that are ringed as chicks and released in the $t$th time period. The multinomial cell probabilities $\mathbf{q}_t \coloneqq \{q_{t,s} \colon t_1 < s \leq t_2+2\}$ are given by
\begin{align}
 q_{t,s} \coloneqq 
 \begin{cases}
   0, & \text{if $t_1 < s \leq t$,}\\
   (1-\phi_{\min\{s-t,\adult\},s-1}) \lambda_{s-1} \prod_{a = 1}^{s-t-1} \phi_{\min\{a,\adult\},t+a-1} , & \text{if $t < s \leq t_2+1$,\!\!}\\
   1-\sum_{s=t_1+1}^{t_2+1} q_{t,s}, & \text{if $s = t_2+2$.}
 \end{cases}
\end{align}

\subsection{Parametrisation}

\subsubsection{Parameters common to all models}

We consider additional covariate information to explain temporal variability. The recovery probabilities are assumed to be logistically regressed on the normalised covariate $\timeCovar_t$ which represent the normalised (bird) year $t$:
\begin{align}
  \logit \lambda_t = \alpha_0 + \beta_0 \timeCovar_t, \quad \text{for $t=t_1,\dotsc,t_2-1$.}
\end{align}
We specify the survival probabilities to be logistically regressed on the normalised covariate $\fdaysCovar_t$ which represents the (normalised) number of days in (bird) year $t$ on which the mean daily temperature fell below freezing in Central England: 
\begin{align}
  \logit \phi_{a,t} = \alpha_a + \beta_a \fdaysCovar_t, \quad \text{for $t=1,\dotsc,T-1$.} \label{eq:herons_survival_probabilities}
\end{align}
Finally, the free parameter in the negative-binomial observation equation is parametrised as $\kappa = \logit^{-1}(\omega) \in (0,1)$ with $\omega \in \reals$.


\subsubsection{Models for the productivity rate}

We specify a set of models for which we perform model comparison on the productivity rates. 
The unknown model parameters are given by $\param = \{\omega,  \alpha_0, \beta_0, \alpha_1, \dotsc, \alpha_\adult, \beta_1, \dotsc, \beta_\adult, \varParam\}$, where $\varParam$ represents the additional model parameters needed for one of the following models for the productivity rate.

\begin{description}
 \item[{Constant productivity.}]
 We set $\log \rho_t = \psi$; $\varParam = \psi$. 

\item[{Productivity regressed on frost days}.]
We set $\log \rho_t = \gamma_0 + \gamma_1 \fdaysCovar_{t-1}$; $\varParam = \{\gamma_0,\gamma_1\}$.

\item[{Direct density dependence}.] 
We set the log-productivity to be a linear function of abundance, $\log \rho_t = \varepsilon_0 + \varepsilon_1 \tilde{y}_t$, where $\tilde{y}_t$ denotes the $t$th normalised observation; 
$\varParam = \{\varepsilon_0,\varepsilon_1\}$. This is one of the models considered by \citet{besbeas2012threshold}.

\item[{Threshold dependence}.]
\citet{besbeas2012threshold} also investigate models in which the productivity is a step function with $K$ levels and hence $K-1$ thresholds ($K$ itself may be unknown) which is defined in terms of the observations. More specifically, the productivity rates are constant between the change-points and monotonically decreasing with increasing population size, \IE assuming that $K > 1$,
  \begin{align}
    \rho_t = 
    \begin{cases}
    \nu_1,   & \text{if $y_t < \tau_1$,}\\
    \nu_{k}, & \text{if $\tau_{k-1} \leq y_t < \tau_k$ for $1 < k < K$,}\\
    \nu_K,   & \text{if $\tau_{K-1} \leq y_t$,}
    \end{cases} \label{eq:productivity_rates_besbeas_morgan}
  \end{align}
  where $\nu_1 > \nu_2 > \dotsb > \nu_{K}$ and $\tau_1 < \tau_2 < \dotsb < \tau_{K-1}$. Thus it is assumed that larger population sizes induce lower productivity rates. For example, this may be due to an exhaustion of high quality breeding sites leading to a reduction in the quantity/quality of young. To ensure these inequalities we specify $\nu_K = \exp(\zeta_K)$ as well as
  \begin{align}
    \nu_k & = \textstyle \sum_{l=k}^K \exp(\zeta_l),\\    
  \tau_k &= \textstyle y_{\min{}} + (y_{\max{}} - y_{\min{}})\dfrac{\smash{\sum_{l=1}^k \exp(\eta_l)}}{\sum_{m=1}^K \exp(\eta_m)},
  \end{align}
  for $k \in \{1,\dotsc,K-1\}$, where $y_{\min{}} = \min\{y_1, \dotsc, y_\nObservationsCount\}$ and $y_{\max{}} = \max\{y_1, \dotsc, y_\nObservationsCount\}$. In this case, $\varParam = \{ \zeta_k, \eta_k: 1\leq  k \leq K\}$.

  
\item[{Regime switching dynamics}.]  
 To constract a more flexible model for the productivity rate, we extend the latent states $\latentVec_t$ by including an additional (unobserved) regime indicator variable $r_t$ which takes values in $\{1,\dotsc,K\}$. Conditional on $r_t$, the productivity rate $\rho_{t-1}$ is then defined as $\rho_{t-1} = \nu_{r_t}$, where $\nu_{1},\dotsc,\nu_K$ are specified as in the threshold model, above. The evolution of the latent regime indicator $r_t$ is assumed to be a Markov chain with transition equation 
  \begin{align}
   r_t|r_{t-1},\param \sim \dMult(K, \mathbf{P}_{r_{t-1}} ),
  \end{align}
  where $\mathbf{P}_{k} = (P_{k,1}, \dotsc, P_{k,K})$ with 
  \begin{align}
    P_{k,l} = \frac{\exp(\varpi_{k,l})}{\sum_{m=1}^K \exp(\varpi_{k,m})}, \quad \text{for $1\leq l\leq K$,}
  \end{align}
  is the $k$th row of the $(K,K)$-transition matrix for the regime indicator variable. In this case, $\varParam = \{\zeta_k,\varpi_{k,l}\colon (k,l)\in \{1,\dotsc,K\}^2\}$.
\end{description}
Finally, we note that we also vary the number of levels, $K$, and the number of age groups, $\adult$, so that the number of models to be compared is much larger than the five specifications for the productivity rate summarised above.

\subsubsection{Prior specification}

We assume that all the model parameters in $\param$ are independent a-priori with $\dN(0,1)$ priors, except that $\omega \sim \dN(-2,4)$. 
 
\subsection{Results}

\begin{figure}[!ht]
 \hspace{-1.5cm}\includegraphics[scale=1]{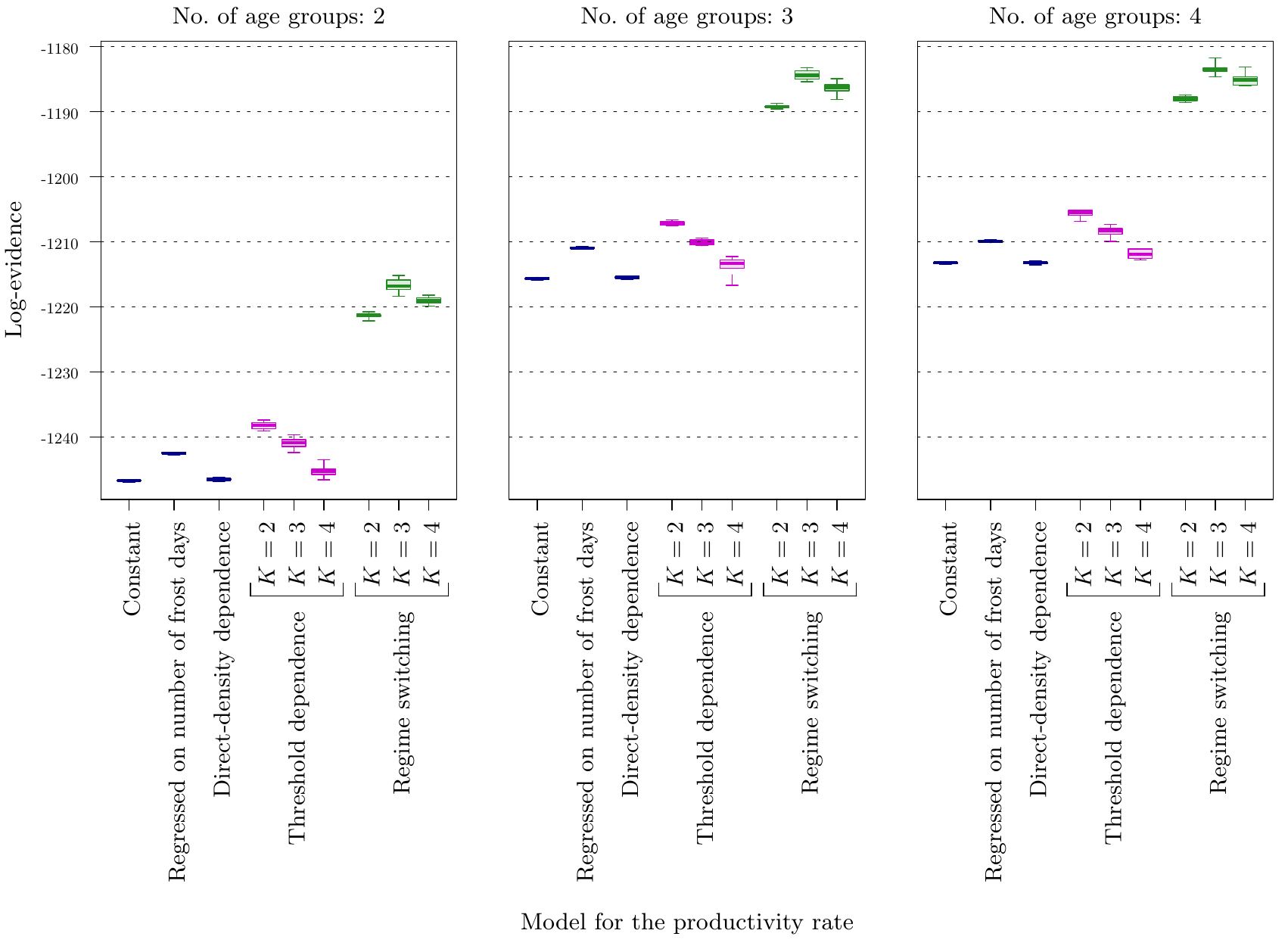}
  \vspace{-0.3cm}
 \caption{Logarithm of estimates of the evidence for different models for the grey herons. Shown are results for the different models for the productivity rate and different numbers of distinctly modelled age categories ($\adult$). For the threshold and regime-switching models, we also investigate different values for the number of thresholds/regimes ($K$). Obtained from $10$ independent runs of the adaptive \gls{SMC} sampler using $1,000$ particles; the \glspl{PF} used to approximate the count-data likelihood employed $4,000$ particles. The average computation time was $42$--$61$ hours for the threshold models,  $32$--$45$ hours for the regime-switching models and $29$--$48$ for the remaining models, the lower numbers corresponding to $\adult=2$ age categories and the higher numbers to $\adult=4$ age categories.}
 \label{fig:herons_evidence_boxplot}
\end{figure}

\begin{figure}[!ht]
 \vspace{-1cm}
  \noindent{}
  \centering
  \begin{subfigure}[t]{\linewidth}
    \centering
    \includegraphics[scale=0.9]{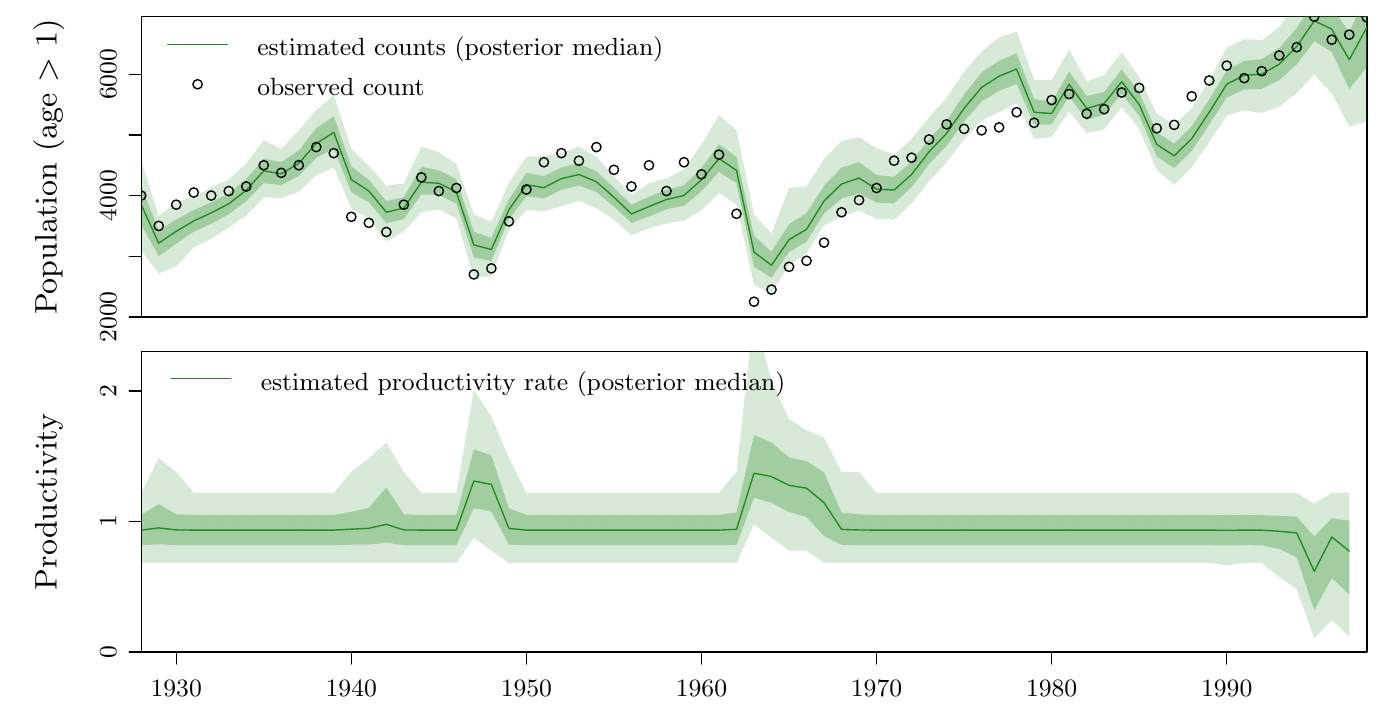}
    \caption{Threshold dependence ($K=4$ levels, \IE $3$ thresholds).} \label{fig:herons_overview_threshold}	
  \end{subfigure}
  \noindent{}
  \centering
  \begin{subfigure}[t]{\linewidth}
    \centering
    \includegraphics[scale=0.9]{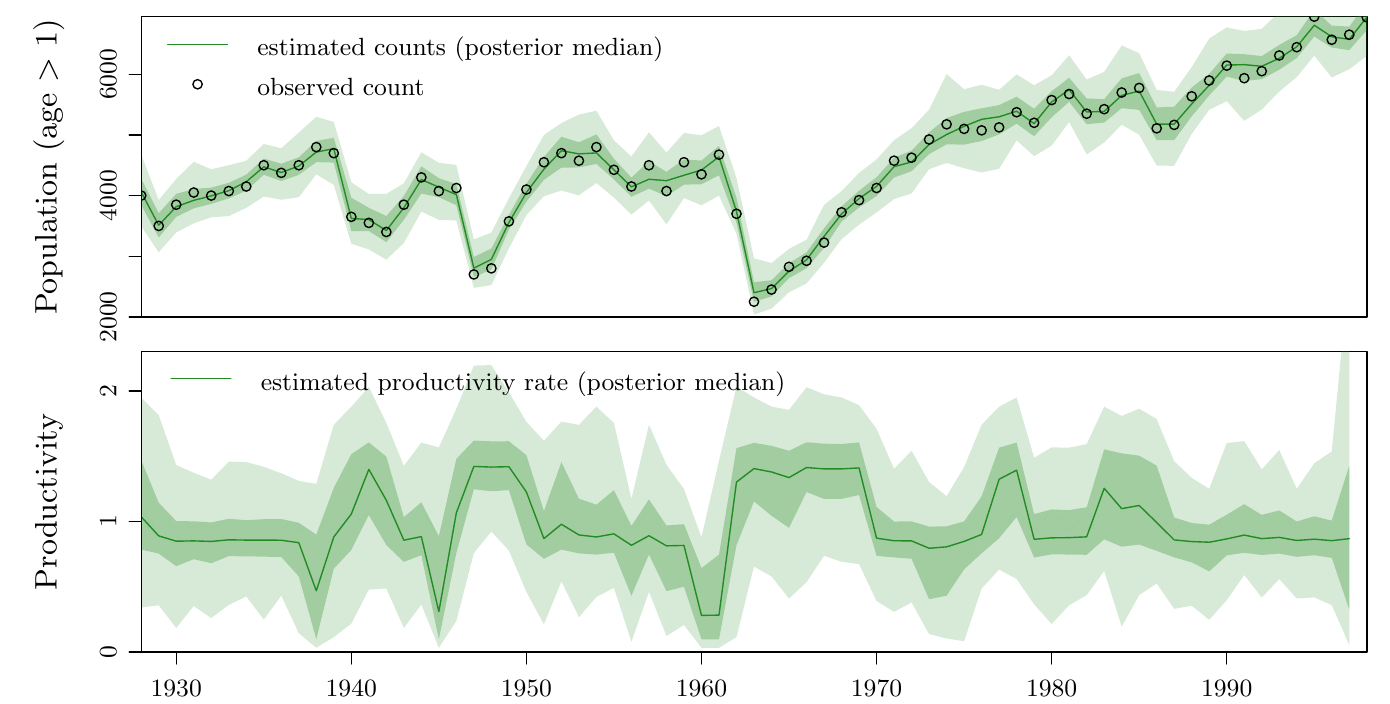} 
    \caption{Regime switching ($K=4$ regimes).}\label{fig:herons_overview_switching}
  \end{subfigure}
\vspace{-0.2cm}

 \caption{Marginal posterior distributions of the estimated heron counts (top rows) and productivity rates (bottom rows) for the threshold model from \citet{besbeas2012threshold} and the novel regime-switching model (results for other models are shown in Web Appendix~B) with $\adult=4$ distinct age categories. The shaded areas represent, respectively, the $90$~\% quantile and range of all encountered realisations. The shown results display the average over $10$ independent repeats of the adaptive \gls{SMC} sampler (each using $1,000$ particles). The \glspl{PF} used to approximate the count-data likelihood use $4,000$ particles.}
  \label{fig:herons_overview}
\end{figure}

Estimates of the evidence for the different models can be found in Figure~\ref{fig:herons_evidence_boxplot}. The fit of the different types of models for the productivity rates is illustrated in Figure~\ref{fig:herons_overview} below (see also Web Figure~2 in Web Appendix~B). Due to the increased flexibility of the productivity rates, the regime-switching model leads to a smaller measurement error. In addition, the evidence for the regime-switching model is much higher than for any of the other models in Figure~\ref{fig:herons_evidence_boxplot}.


Figure~\ref{fig:herons_evidence_boxplot} supports the finding from \citet{besbeas2009completing} that modelling the herons using four age groups is appropriate (though the results with three age groups are similar). However,  using only two age groups drastically reduces the model evidence across all specifications for the productivity rate. The results also support the findings from \citet{besbeas2012threshold} that the first three models (with productivity rate constant, regressed on the number of frost days, or density-dependent) do not explain the data well.

The posterior distribution of the productivity rate (under any of the models) must be interpreted with care. Indeed, note the sharp decline of the productivity rate in the years immediately preceding the severe winters of 1946--47 and 1962--63 in Figure~\ref{fig:herons_overview_switching}. This indicates that the linear model for the survival rates in \eqref{eq:herons_survival_probabilities} may not be flexible enough to accommodate the drop of the heron population in subsequent years.
 
We also implemented all of the above-mentioned models using a continuous (linear-Gaussian) approximation to the state-space model for the count data. For the regime-switching model, a \gls{PF} is then still necessary to sample the latent regime indicators. However, as these take values in a small finite set, this can be done highly efficiently using the so called \emph{discrete} \gls{PF} from \citet{fearnhead2003online}. The results (omitted here) are relatively similar to the results obtained for the original models, \IE the approximation did not affect the ordering of the models in terms of their evidence. However, the regularising effect of the continuous (linear-Gaussian) approximation artificially increased the evidence for all models by roughly the same amount. In other words, such linear-Gaussian approximations (employed for computational reasons) lead to an overestimation of the model fit.

\section{Conclusion}
\label{sec:conclusion}

We have proposed an efficient Monte Carlo methodology for Bayesian parameter estimation and model comparison for integrated population models which have a state-space model for the noisily observed population sizes as one of their constituent parts. Utilising \gls{PMCMC} techniques, our approach can be generally applied to such models, requiring neither (a) approximate linear or Gaussian modelling assumptions which introduce a bias that is often difficult to quantify nor (b) data-augmentation schemes which can lead to poor mixing in \gls{MCMC} algorithms if highly correlated states or parameters are updated separately. Incorporating these ideas into an \gls{SMC} sampler also yields estimates of posterior model probabilities relating to the dependence of the model parameters, including even the number of age groups for the true states. Finally, we have proposed two extensions which enhance the efficiency of our methodology by exploiting the structure of integrated population models.


We have demonstrated the methodology on two different applications: (1) little owls and (2) grey herons. For the owls, we found no evidence in favour of some of the more complex model specifications proposed in the literature, \EG for the dependence of immigration on the abundance of voles proposed in \citet{abadi2010estimation}. For the herons, we showed that existing models, including the state-of-the-art threshold model for the productivity from \citet{besbeas2012threshold}, do not explain the data well. To remedy this problem, we proposed a novel regime-switching model and demontrated that it is very strongly favoured over the other models in terms the Bayes factor. Our methodology is related to the \gls{SMC}\textsuperscript{2} algorithm from \citet{drovandi2016alive}. However, even in low-dimensional settings (\IE $3$-$4$ unknown model parameters) \citet{drovandi2016alive} 
had to combine \gls{SMC}\textsuperscript{2} with another importance-sampling algorithm to obtain evidence estimates accurate enough for model comparison in some examples (and, as pointed out by \citet{drovandi2016alive}, this importance sampling scheme may not be applicable in higher dimensions). In contrast, in all applications considered in this work, the evidence estimates provided by our methodology were accurate enough to directly identify the best-performing models despite the relatively large number of unknown model parameters (\IE $6$--$58$ for the owls; $8$--$31$ for the herons).

\section*{Acknowledgements}

A.B.\ and A.F.\ were funded by a Leverhulme Trust Prize. R.K.\ and P.D.\ were part funded by the The Alan Turing Institute under the Engineering and Physical Sciences Research Council grant EP/N510129/1. \vspace*{-8pt}

\section*{Supplementary Materials}

Web Appendices and Figures, referenced in Sections~\ref{sec:parameter_estimation}--\ref{sec:herons}, are available with this paper. 


\renewcommand*{\bibfont}{\footnotesize}
\setlength{\bibsep}{3pt plus 0.3ex}
\bibliography{joint}


\clearpage
\appendix

\pagenumbering{roman}
\setcounter{page}{1}
\setcounter{section}{0}
\setcounter{subsection}{0}
\setcounter{figure}{0}
\setcounter{equation}{0}
\setcounter{table}{0}

\renewcommand\thesection{\Alph{section}}
\renewcommand\thesubsection{\thesection.\arabic{subsection}}

\renewcommand\thetable{\thesection.\arabic{table}}
\renewcommand\thefigure{\thesection.\arabic{figure}}
\renewcommand\theequation{\thesection.\arabic{equation}}

\begin{center}
 \LARGE{Web-based Supplementary Materials for `Efficient sequential Monte Carlo algorithms for integrated population models' by A.~Finke, R.~King, A.~Beskos and P.~Dellaportas}
\end{center}

\section{Further details on the owls example}
\label{sup:sec:owls}

\subsection{Evidence estimates}
\label{sup:subsec:owls_evidence}
  
\begin{figure}[!ht]
 \centering
 \includegraphics[scale=1]{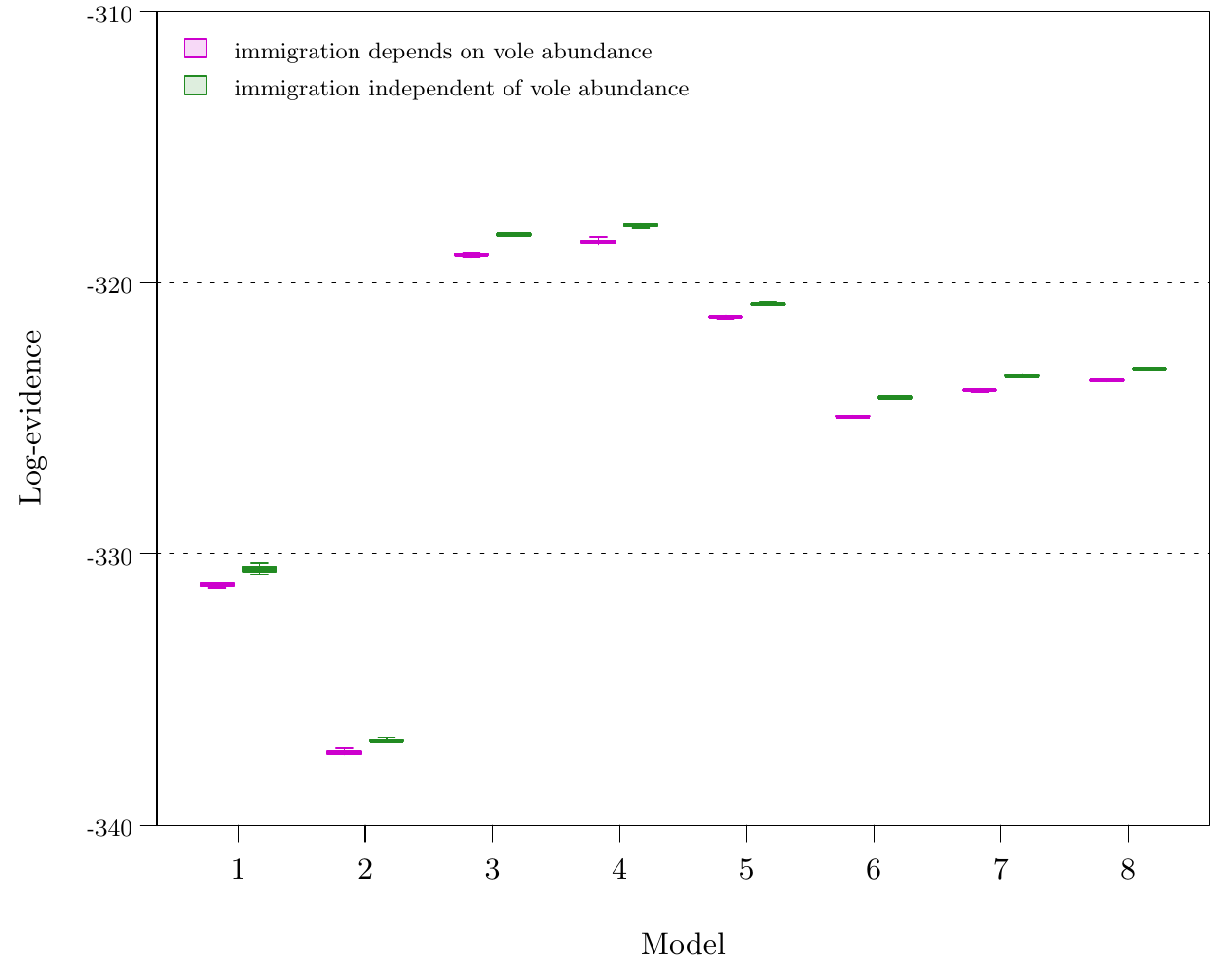}
\vspace{-0.2cm}
 \caption{Logarithm of the estimates of the evidence for the eight models for the little owls with or without dependence of the immigration rate on the abundance of voles. The results were obtained from $20$ independent runs of the adaptive \gls{SMC} sampler using $10,000$ particles; the \glsdescplural{PF} used to approximate the marginal likelihoods use $1,000$ particles. The average computation time for each \gls{SMC} sampler was around $9$--$18$ hours on a single core (we stress such a relatively large number of particles was only used to gain accurate evidence estimates in the large models (in terms of the number of parameters), \IE in Models $1$--$5$; for the smaller models, \IE Models~$6$--$8$, quite similar results could have been obtained in $30$ minutes by using only $500$ particles.}
  \label{fig:owls_evidence_boxplot}
\end{figure}

Figure~\ref{fig:owls_evidence_boxplot} shows estimates of the evidence for the following eight models. For each of these, we consider the case that the immigration rate depends on the abundance of voles as in \citet{abadi2010estimation} (\IE we estimate $\delta_1 \neq 0$) and the case that the immigration rate does not depend on the abundance of voles (\IE we enforce $\delta_1 = 0$). 

\begin{enumerate}

 \item \label{enum:owls:model:1} full model as specified in Section~\ref{sec:owls} (with the Gaussian priors specified there) -- \IE both the productivity rates $\rho_t$ and the recapture probabilities $p_{a,t}$ are time-varying. 
 
 \item \label{enum:owls:model:2} like Model~\ref{enum:owls:model:1} but with constant capture probabilities over time, \IE enforcing $p_{a,2} = \dotsc = p_{a,T}$, for all $a \in \{\juvenile, \adult\}$ by setting $\beta_{2} = \dotsc = \beta_{\nObservationsCount} = \beta$. 
 
 \item \label{enum:owls:model:3} like Model~\ref{enum:owls:model:1} but with constant productivity rates over time, \IE enforcing $\rho_1 = \dotsc = \rho_T$ by setting $\gamma_1 = \dotsc = \gamma_T = \gamma$. 
 
 \item \label{enum:owls:model:4} like Model~\ref{enum:owls:model:3} but additionally assuming that the survival probabilities do not follow a  linear trend, \IE enforcing $\alpha_3 = 0$. 
 
 \item \label{enum:owls:model:5} like Model~\ref{enum:owls:model:4} but additionally assuming that survival probabilities do not depend on the gender, \IE enforcing $\alpha_1 = 0$.

 \item \label{enum:owls:model:6} like Model~\ref{enum:owls:model:2} but with constant productivity rates over time, \IE enforcing $\rho_1 = \dotsc = \rho_T$ by setting $\gamma_1 = \dotsc = \gamma_T = \gamma$. 
 
 \item \label{enum:owls:model:7} like Model~\ref{enum:owls:model:6} but additionally assuming that the survival probabilities do not follow a  linear trend, \IE enforcing $\alpha_3 = 0$. 
 
 \item \label{enum:owls:model:8} like Model~\ref{enum:owls:model:7} but additionally assuming that survival probabilities do not depend on the gender, \IE enforcing $\alpha_1 = 0$.
\end{enumerate}

While there is little difference in the evidence for Models~\ref{enum:owls:model:3} and \ref{enum:owls:model:4}, Figure~\ref{fig:owls_evidence_boxplot} makes it clear that the data do not support the hypothesis that the immigration rate depends on the abundance of voles. 

\subsection{Efficiency of the refined tempering scheme}
\label{sup:subsec:owls_efficiency}

In Table~\ref{tab:double_tempering_advantage}, we illustrate efficiency gains attainable through our refined likelihood tempering scheme (proposed in Section~4.3 of the main manuscript) over standard likelihood tempering scheme from \citet{duan2015density} (see Section~4.2 of the main manuscript, for details). That is, for each of the eight models specified in Subsection~\ref{sup:subsec:owls_evidence} above, Table~\ref{tab:double_tempering_advantage} displays
\begin{equation}
 (\textit{efficiency gain}) =  \frac{\mathit{MSE} \times (\textit{computation time})}{\mathit{MSE} \times (\textit{computation time})} \myatop{\}\,\text{\footnotesize{standard tempering}}}{\}\,\text{\footnotesize{refined tempering}}} . \label{eq:efficiency_gain}
\end{equation}
Here, $\mathit{MSE}$ denotes the average \gls{MSE} of the estimate of the posterior mean\footnote{Results for other estimates of interest (\EG the posterior variance) were similar as those for the posterior means and are therefore suppressed.} based on $20$ independent repeats of the \gls{SMC} samplers (the average is taken over all components of the vector of model parameters). Since the true posterior means are intractable, we ran an \gls{MCMC} algorithm using a large number ($10,000,000$) of iterations for each model and treated the resulting posterior mean estimates as the true values. Furthermore, $(\textit{computation time})$ represents the average computation time over the independent repeats.  

Table~\ref{tab:double_tempering_advantage} displays the efficiency gains obtained through the refined tempering scheme using different numbers of particles. To simplify the presentation, we only show results for the case that $\delta_1 \neq 0$, \IE we allow for dependence of immigration on the abundance of voles.

\begin{table}
 \centering
  \caption{Average efficiency gain (as defined in \eqref{eq:efficiency_gain}) of the refined likelihood tempering scheme (see Section~4.3 of the main manuscript) over standard likelihood tempering for different numbers of particles ($\nParticlesUpper$).}
  \begin{tabular}{ld{2.1}d{2.1}d{2.1}d{2.1}d{2.1}d{2.1}d{2.1}d{2.1}}
    \hline
    Model & \myalign{c}{1} & \myalign{c}{2}& \myalign{c}{3} & \myalign{c}{4} & \myalign{c}{5} & \myalign{c}{6} & \myalign{c}{7} & \myalign{c}{8} \\ \hline
    \textit{efficiency gain} ($\nParticlesUpper = 1,000$) & 14.0 & 4.7 & 2.8 & 2.3 & 2.9 & 0.9 & 0.9 & 1.2\\
    \textit{efficiency gain} ($\nParticlesUpper = 10,000$)  & 19.1 & 4.5 & 2.3 & 2.3 & 2.5 & 0.8 & 1.2 & 0.6\\
    \hline
  \end{tabular}
   \label{tab:double_tempering_advantage}
\end{table}

\section{Further details on the herons example}
\label{sup:sec:herons}

\begin{figure}[!ht]
  \noindent{}
  \centering
  \begin{subfigure}[t]{\linewidth}
    \centering
    \includegraphics[scale=0.8]{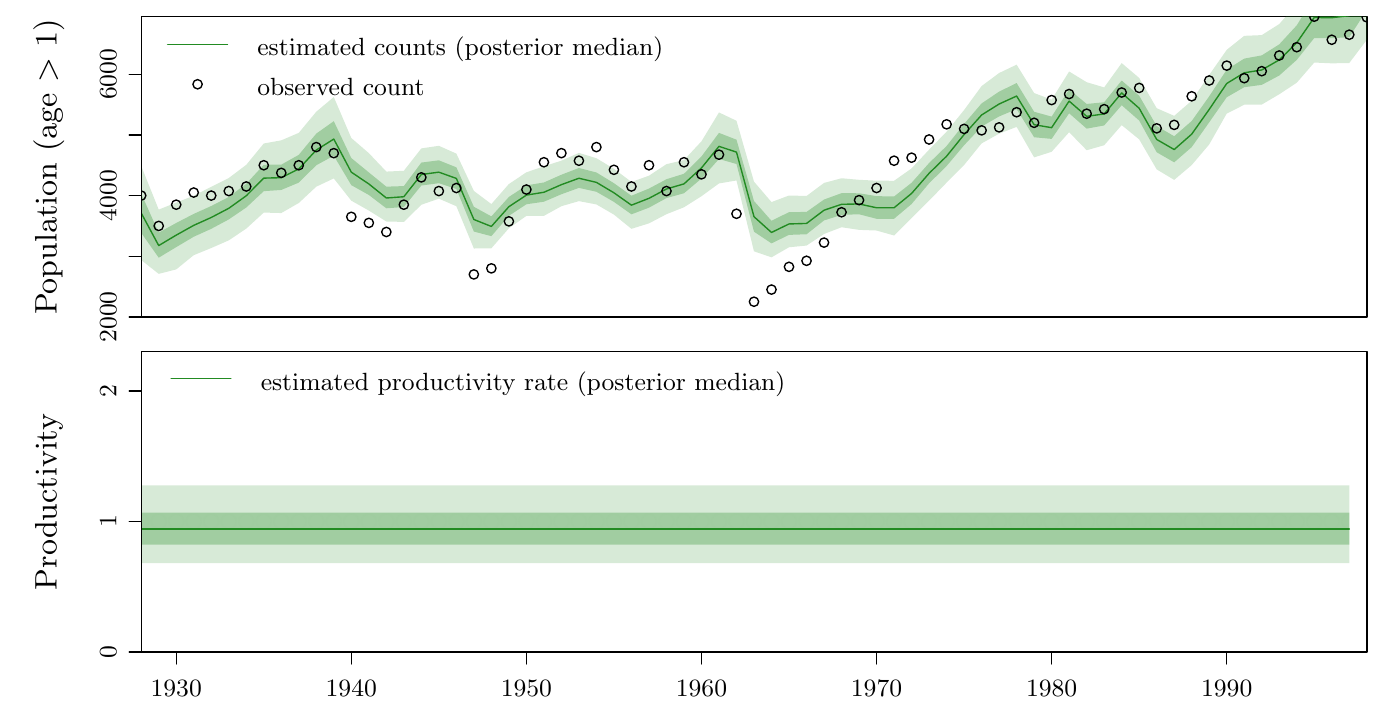}
    \caption{Constant productivity.} \label{fig:herons_overview_constant}	
  \end{subfigure}
  \noindent{}
  \centering
  \begin{subfigure}[t]{\linewidth}
    \centering
    \includegraphics[scale=0.8]{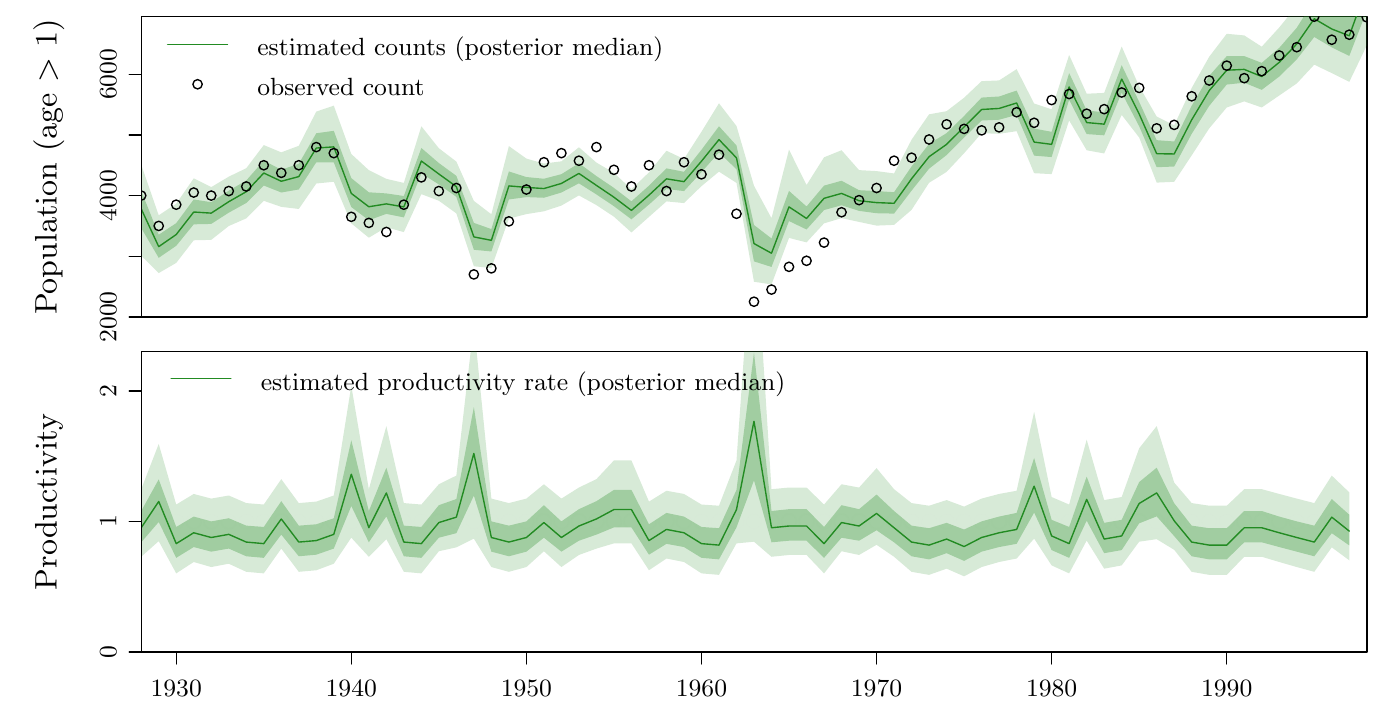}
    \caption{Productivity regressed on frost days.} \label{fig:herons_overview_fdays}	
  \end{subfigure}
  \noindent{}
  \centering
  \begin{subfigure}[t]{\linewidth}
    \centering
    \includegraphics[scale=0.8]{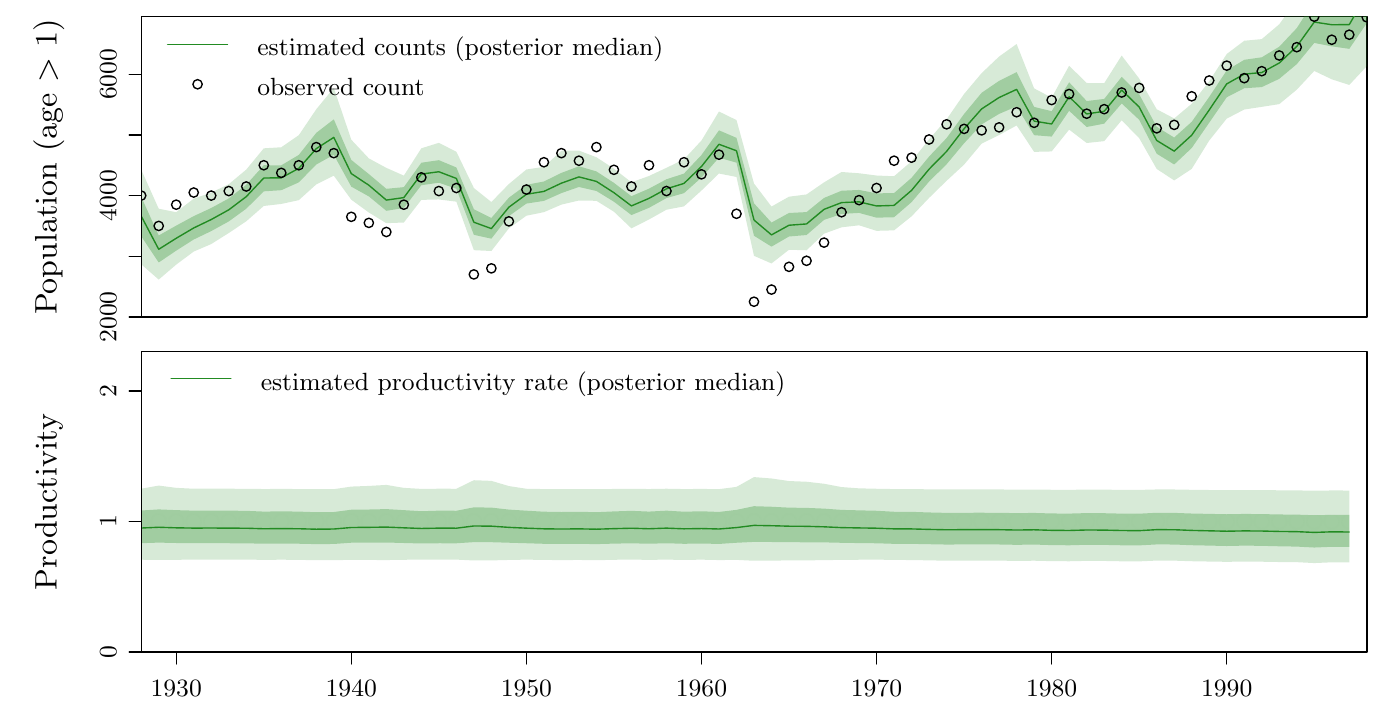} 
    \caption{Direct density dependence.}\label{fig:herons_overview_density}	
  \end{subfigure}
\vspace{-0.2cm}
 \caption{Marginal posterior distributions of the estimated heron counts and productivity rates in the same setting as in Figure~3 in the main manuscript.}
  \label{fig:herons_overview_1}
\end{figure}

Figure~\ref{fig:herons_overview_1} shows the marginal posterior distributions of the estimated heron counts and productivity rates for some of the models not included in Figure~3 in the main manuscript.

\enlargethispage{4\baselineskip}
\label{sup:sec:herons}

\section{Detailed algorithms}
\label{sup:sec:extensions}

\subsection{Full particle filter}
\label{sup:subsec:full_pf}

In this section, we state slightly extended versions of the \gls{SMC} algorithm for filtering (Algorithm~1 in the main manuscript) and of the \gls{SMC} algorithm for model comparison (Algorithm~4 in the main manuscript). 

We begin by stating (in Algorithm~\ref{alg:full_pf}) a version the \gls{SMC} algorithm for filtering employed within our simulations. It employs a low-variance resampling scheme termed \emph{systematic} resampling \citep{carpenter1999improved} and which resamples only when the \gls{ESS} falls below some pre-specified threshold $\ESS^\star \in (0,1)$. For our simulations, we used $\ESS^\star = 0.9$. \citep{delmoral2012adaptive}. Since we do not necessarily resample at every step, the estimator $\hat{p}(\dataCount|\param)$ provided by Algorithm~\ref{alg:full_pf} differs slightly from the estimator defined in Section~3.1 of the main manuscript (see \EG\citep[Equation~15]{delmoral2006sequential}. Note also that in Algorithm~\ref{alg:full_pf}, the \gls{ESS} is scaled such that it takes values between $0$ and $1$. This potentially differs from some of the literature in which this quantity is scaled such that it takes values between $1$ and $\nParticlesLower$. Finally, note that Algorithm~\ref{alg:full_pf} permits the use of proposal kernels $\proposalKernelLower_{\param, t}(\latentVec_t|\latentVec_{t-1})$ which differ from the model transitions $f_{\param}(\latentVec_t|\latentVec_{t-1})$ \citep{doucet2000sequential}. This could be even further generalised to the use of  auxiliary \glspl{PF} \citep{pitt1999filtering, johansen2008note}.

As always, Algorithm~\ref{alg:full_pf} uses the convention that actions prescribed for the $n$th particle are to be performed \emph{conditionally independently} for all $n \in \{1, \dotsc, \nParticlesLower\}$. 

\noindent\parbox{\textwidth}{
\begin{flushleft}
  \begin{framedAlgorithm}[PF with adaptive systematic resampling] \label{alg:full_pf}~ \!\!\!
    \begin{flushleft}
      \begin{enumerate}
	\item At Step~$1$,
	\begin{enumerate}
	  \item sample $\smash{\latentVec_1^n \sim \proposalKernelLower_{\param, 1}}(x_1)$,
	  \item set $\smash{\WeightLower_1^n \coloneqq w_1^n / \sum_{k=1}^\nParticlesLower \weightLower_1^k}$, where $\smash{\weightLower_1^n \coloneqq \dfrac{\mu_\param(\latentVec_1^n)g_\param(y_1|\latentVec_1^n)}{\proposalKernelLower_{\param, 1}(\latentVec_1^n)}}$.
	\end{enumerate}
	\item At Step~$t = 2,\dotsc,\nStepsLower$, 
	\begin{enumerate}
	  \item \textbf{If} $\ESS_{t-1} \coloneqq 1/\sum_{n=1}^\nParticlesLower (\WeightLower_{t-1}^n)^2 < \ESS^\star$,
	  \begin{enumerate}
	    \item $\xi_{t-1} \coloneqq 1$,
	    \item sample $\smash{\ancestorIndexLower_{t-1}^{1:\nParticlesLower}}$ via systematic resampling based on the weights $\WeightLower_{t-1}^{1:\nParticlesLower}$.
	  \end{enumerate}
	  \textbf{Else,} 
	  \begin{enumerate}
	    \item $\xi_{t-1} \coloneqq 0$,
	    \item set $\smash{\ancestorIndexLower_{t-1}^n \coloneqq n}$.
	  \end{enumerate}
	  \item sample $\smash{\latentVec_t^n \sim \proposalKernelLower_{\param, t}(x_t|\latentVec_{t-1}^{a_{t-1}^n})}$,
	  \item set $\smash{\WeightLower_t^n \coloneqq \weightLower_t^n / \sum_{k=1}^\nParticlesLower \weightLower_t^k}$, where 
	  \begin{align}
	   \weightLower_t^n \coloneqq \bigl[\ind\{\xi_{t-1} = 1\} + \weightLower_{t-1}^{n}\ind\{\xi_{t-1} = 0\}\bigr] \smash{\dfrac{f_\param(\latentVec_t^n|\latentVec_{t-1}^{\ancestorIndexLower_{t-1}^n})g_\param(y_t|\latentVec_t^n)}{\proposalKernelLower_{\param, t}(\latentVec_t^n|\latentVec_{t-1}^{\ancestorIndexLower_{t-1}^n})}}.
	  \end{align}
	\end{enumerate}
	\item Set $\mathcal{T} \coloneqq \{t \in \{1,\dotsc,\nStepsLower-1\} \mid \xi_t = 1\} \cup \{\nStepsLower\}$ and
	\begin{equation}
	  \hat{p}(\dataCount|\param) \coloneqq \prod_{t \in \mathcal{T}} \frac{1}{\nParticlesLower}\sum_{n=1}^\nParticlesLower \weightLower_t^n.
	\end{equation}
      \end{enumerate}
    \end{flushleft}
  \end{framedAlgorithm}
\end{flushleft}
}

\subsection{Full SMC algorithm for evidence approximation}
\label{sup:subsec:full_smc}

In this subsection, we state the \gls{SMC} sampler which we employ to perform inference in the applications described in Sections~\ref{sec:owls} and \ref{sec:herons}. More precisely, we now employ the \gls{PF} with adaptive systematic resampling from Algorithm~\ref{alg:full_pf} (instead of Algorithm~\ref{alg:pf}) to approximate the marginal count-data likelihood. Furthermore, we make the following modifications to the \gls{SMC} sampler itself.

\begin{description}
 \item[Adaptive systematic resampling.] We employ systematic instead of multinomial resampling and we resample only when the \gls{ESS} falls below some pre-specified threshold $\ESS^\star \in (0,1)$. For our simulations, we used $\ESS^\star = 0.9$.
 
 \item[Adaptive tempering schedule.] we specify the temperature schedule adaptively in such a way that the \gls{CESS} decreases to a pre-specified amount $\CESS^\star \in (0,1)$ at each iteration \citep{zhou2016towards}. When resampling at every step, this reduces to the \gls{ESS}-based approach introduced in \citet{jasra2010inference} and for which some theoretical justification is provided in \citet{beskos2016convergence}. In this work, we employ two different thresholds, indicated by subscripts $\alpha$ and $\beta$, for tempering the different likelihood terms as described in Section~4.3 of the main manuscript. This provides an easy way to control the number of steps and hence the computational cost allocated to each of the two stages of the algorithm.  In the applications, we used $\CESS_\alpha^\star = 0.9999$ and $\CESS_\beta^\star = 0.99$ for the little owls models\footnote{For the standard tempering scheme needed in the efficiency comparison in Subsection~\ref{sup:subsec:owls_efficiency} above, we used $\CESS^\star = 0.9993$ as this led to a similar computational cost as the refined tempering scheme (nonetheless, recall that the comparison in Subsection~\ref{sup:subsec:owls_efficiency} takes computational cost into account).} and $\CESS_\alpha^\star = \CESS_\beta^\star = 0.9999$ for the grey herons models.

 \item[Adaptive proposal kernel.] We use the particles from previous steps to guide the construction of the proposal distributions for the parameters (\EG see \citet{chopin2002sequential, peters2010ecological}). More specifically, at each step, we propose a new set of parameter values $\varParam$ from the two-component Gaussian mixture
 \begin{equation}
  q(\varParam|\param; \lambda \varSigma) \coloneqq 0.95 \dN(\varParam; \param, 2.38^2 d_\param^{-1} \lambda \varSigma) + 0.05 \dN(\varParam; \param, 0.1^2 d_\param^{-1} \mathbf{I}_{d_\param}).
 \end{equation}
 Here, $d_\param \in \naturals$ denotes the dimension (length) of the parameter vector $\param$ and $\mathbf{I}_n$ is the $(n,n)$-identity matrix. In addition, $\lambda > 0$ is some scaling factor which we adaptively set to keep the acceptance rate of the \gls{MH} updates within reasonable bounds \citep{jasra2010inference}. That is, $\lambda$ is increased by a factor of $2$ (decrease by a factor of $1/2$) if the acceptance rate of the \gls{MH} updates exceeded $0.5$ (dropped below $0.2$). For further tuning of \gls{PMCMC} kernels and the \glsdesc{DA} technique, see \citet{sherlock2015efficiencyHapalike,sherlock2015efficiencyPseudo,doucet2015efficient}.
 
\end{description}

We stress that adapting the tempering schedule or the proposal scale in the way described above no longer guarantees that the evidence estimate is unbiased. However, any potential bias is normally far outweighed by the variance reductions brought about by these modifications.

The resulting \gls{SMC} sampler is outlined in Algorithm~\ref{alg:full_smc_sampler}, where we use the convention that any action specified for \emph{some} Index~$m$ is to be performed conditionally independently for \emph{all} $m \in \{1,\dotsc,\nParticlesUpper\}$. This algorithm is a special case of \citet[Algorithm~4]{zhou2016towards}. Again, we note that the \gls{ESS} and \gls{CESS} are scaled such that they take values between $0$ and $1$. 

\begin{flushleft}
\begin{framedAlgorithm}[adaptive, delayed-acceptance SMC for model choice]~ \!\!\!\label{alg:full_smc_sampler}
\begin{enumerate}
 \item Initialisation, Stage~1:
 \begin{enumerate}
   \item Sample $\smash{\param_0^m  \sim \paramPrior(\param|\modelIndicator_i)}$,
   \item Set $\smash{\widetilde{\WeightUpper}_0^m \coloneqq 1/\nParticlesUpper}$, $\smash{\tilde{\weightUpper}_0^m \coloneqq 1}$ and $\xi_0 \coloneqq 0$, $a_0 \coloneqq 0$, $\lambda \coloneqq 1$ and $s\coloneqq 1$. 
 \end{enumerate}

 \item \label{alg:full_smc_sampler:stage_1} While $\alpha_{s-1} < 1$, writing $\smash{u_{s-1}^m \coloneqq p(\dataOther|\param_{s-1}^m, \modelIndicator_i)}$, do:
 \begin{enumerate}

  \item \label{alg:full_smc_sampler:stage_1:adapt_temperature} set $\alpha_s \coloneqq \alpha \wedge 1$, where $\invTemp$ solves $\CESS_{s-1}(\alpha) = \CESS_\alpha^\star$, where 
  \begin{align}
   \CESS_{s-1}(\alpha) \coloneqq \nParticlesUpper\frac{[\sum_{m=1}^\nParticlesUpper \widetilde{\WeightUpper}_{s-1}^m (u_{s-1}^m)^{\alpha - \alpha_{s-1}}]^2}{\sum_{m=1}^\nParticlesUpper \widetilde{\WeightUpper}_{s-1}^m [(u_{s-1}^m)^{\alpha - \alpha_{s-1}}]^{2}}.
  \end{align}
  
  \item Set $\smash{\WeightUpper_s^m \coloneqq {\weightUpper_s^m}/{\sum_{k=1}^\nParticlesUpper \weightUpper_s^k}}$, where $\smash{\weightUpper_s^m  \coloneqq \tilde{v}_{s-1}^m (u_{s-1}^m)^{\invTemp_s-\invTemp_{s-1}}}$.

  \item Set 
   $\smash{\varSigma_s \coloneqq \sum_{m=1}^\nParticlesUpper \WeightUpper_s^m (\param_{s-1}^m - \mu_s) (\param_{s-1}^m-\mu_s)^\T}$.
  
  \item  \label{alg:full_smc_sampler:stage_1:resample} \textbf{If} $\ESS_s \coloneqq 1/\sum_{m=1}^\nParticlesUpper (\WeightUpper_s^m)^2 < \ESS^\star$, 
  \begin{enumerate}
   \item set $\xi_s \coloneqq 1$, $\tilde{\weightUpper}_s^m \coloneqq 1$ and $\smash{\widetilde{\WeightUpper}_s^m \coloneqq 1/\nParticlesUpper}$.
   \item sample $\smash{\ancestorIndexUpper_{s-1}^{1:\nParticlesUpper}}$ via systematic resampling based on the weights $\WeightUpper_s^{1:\nParticlesUpper}$.
  \end{enumerate}
  \textbf{Else,}
  \begin{enumerate}
    \item set $\xi_s \coloneqq 0$, $\tilde{\weightUpper}_s^m \coloneqq v_{s}^m$ and $\smash{\widetilde{\WeightUpper}_s^m \coloneqq {\tilde{\weightUpper}_s^m}/{\sum_{k=1}^\nParticlesUpper \tilde{\weightUpper}_s^k}}$
    \item set $\smash{\ancestorIndexUpper_{s-1}^{m} \coloneqq m}$,
  \end{enumerate}

  \item \label{alg:full_smc_sampler:stage_1:mcmc_update} Propose $\varParam \sim q(\varParam|\param) \coloneqq q(\varParam|\param; \lambda \varSigma_s)$, where we write $\smash{\param \coloneqq \param_{s-1}^{\ancestorIndexUpper_{s-1}^m}}$, and set $\param_s^m \coloneqq \varParam$ \gls{WP} $\min\{1,r\}$, where
  \begin{equation}
   r \coloneqq \frac{q(\param|\varParam)}{q(\varParam|\param)} \frac{p(\varParam|\modelIndicator_i)}{p(\param|\modelIndicator_i)} \biggl[\frac{p(\dataOther|\varParam, \modelIndicator_i)}{ p(\dataOther|\param, \modelIndicator_i)}\biggr]^{\mathrlap{\alpha_s}};
  \end{equation}
  otherwise, set $\param_s^m \coloneqq \param$.
  
  
  \item (potentially) adapt $\lambda$ and set $s \leftarrow s + 1$.
 \end{enumerate}

 \item Initialisation, Stage~2:
  \begin{enumerate}
   \item Sample $\hat{p}_{s-1}^m(\dataCount|\param_{s-1}^m,\modelIndicator_i)$ using Alg.~\ref{alg:full_pf} (with $\param = \param_{s-1}^m$).
   \item Set $\beta_{s-1} \coloneqq 0$.
 \end{enumerate}
 
 \item While $\beta_{s-1} < 1$, now writing $\smash{u_{s-1}^m \coloneqq \hat{p}_{s-1}^m(\dataCount|\param_{s-1}^m, \modelIndicator_i)p(\dataOther|\param_{s-1}^m, \modelIndicator_i)}$, do:
 \begin{enumerate}
  \item Perform Steps~\ref{alg:full_smc_sampler:stage_1:adapt_temperature}--\ref{alg:full_smc_sampler:stage_1:resample} (with $\beta_s$ and $\CESS_\beta^\star$ instead of $\alpha_s$ and $\CESS_\alpha^\star$).
  
  \item Sample $\param_s^m$ via Algorithm~\ref{alg:pmmh_delayed_acceptance} (using Alg.~\ref{alg:full_pf} instead of Alg.~\ref{alg:pf} in to approximate the marginal likelihood in Step~\ref{alg:pmmh_delayed_acceptance:3} (and with $\alpha = \beta_{s}$, $\param=\param_{s-1}^{\ancestorIndexUpper_{s-1}^m}$, $\hat{p}(\dataCount|\param)=\hat{p}_{s-1}^{\ancestorIndexUpper_{s-1}^m}(\dataCount|\param_{s-1}^{\ancestorIndexUpper_{s-1}^m},\modelIndicator_i)$ and $q(\varParam|\param) = q(\varParam|\param; \lambda \varSigma_s)$).
  
  \item (potentially) adapt $\lambda$ and set $s \leftarrow s + 1$.
 \end{enumerate}
\item Set $\nStepsUpper \coloneqq s-1$, $\mathcal{S} \coloneqq \{s \in \{1,\dotsc,\nStepsUpper-1\}\mid \xi_s = 1\} \cup \{\nStepsUpper\}$ and
\begin{equation}
\hat{p}(\data|\modelIndicator_i) \coloneqq \prod_{s \in \mathcal{S}} \frac{1}{\nParticlesUpper}\sum_{m=1}^\nParticlesUpper \weightUpper_s^m.
\end{equation}
\end{enumerate}
\end{framedAlgorithm}
\end{flushleft}



\end{document}

%% file: mathematics.tex
\newcommand*{\mycmss}{\fontfamily{cmss}\selectfont}
\DeclareTextFontCommand{\textcmss}{\mycmss}

\newcommand{\spaceTheta}{\textup{\text{\mycmss{\greektext J\latintext}}}}

\newcommand{\nParticlesUpper}{M} 
\newcommand{\nStepsUpper}{S} 
\newcommand{\nStepsUpperFirst}{S'} 
\newcommand{\extendedTarget}{\pi} 
\newcommand{\testFun}{\varphi} 
\newcommand{\invTemp}{\alpha} 
\newcommand{\invTempFirst}{\alpha} 
\newcommand{\invTempSecond}{\beta} 
\newcommand{\ancestorIndexUpper}{b} 
\newcommand{\weightUpper}{v} 
\newcommand{\WeightUpper}{V} 

\newcommand{\proposalKernelLower}{q} 
\newcommand{\nParticlesLower}{N} 
\newcommand{\nStepsLower}{T} 
\newcommand{\ancestorIndexLower}{a} 
\newcommand{\weightLower}{w} 
\newcommand{\WeightLower}{W} 

\newcommand{\modelIndicator}{\mathcal{M}} 
\newcommand{\modelSet}{\mathcal{I}} 

\newcommand{\param}{{\bmath{\theta}}} 

\newcommand{\varParam}{{\bmath{\vartheta}}} 

\newcommand{\paramSet}{\spaceTheta} 
\newcommand{\paramPrior}{p} 

\newcommand{\latentVec}{\mathbf{x}} 
\newcommand{\latent}{x} 
\newcommand{\latentFull}{\mathbf{x}} 
\newcommand{\latentSet}{\mathsf{X}} 

\newcommand{\data}{\mathbf{z}} 
\newcommand{\dataCount}{\mathbf{y}} 
\newcommand{\dataRing}{\mathbf{m}} 
\newcommand{\dataRecapture}{\mathbf{m}} 
\newcommand{\dataFecundity}{\mathbf{n}} 
\newcommand{\dataOther}{\mathbf{w}} 

\newcommand{\nObservationsCount}{T} 

\newcommand{\fdaysCovar}{\mathit{fdays}} 
\newcommand{\timeCovar}{\mathit{time}} 
\newcommand{\yearCovar}{\mathit{year}} 
\newcommand{\voleCovar}{\mathit{vole}} 

\DeclareMathOperator{\dN}{Normal} 
\DeclareMathOperator{\dPois}{Poisson} 
\DeclareMathOperator{\dMult}{Multinomial} 
\DeclareMathOperator{\dBin}{Binomial} 
\DeclareMathOperator{\dNegBin}{Negative-Binomial} 

\newcommand{\diff}{\mathrm{d}} 
\newcommand{\intDiff}{\,\mathrm{d}} 
\DeclareMathOperator{\E}{\mathbb{E}} 
\DeclareMathOperator{\var}{var} 
\DeclareMathOperator{\ind}{\mathbb{I}} 
\DeclareMathOperator{\logit}{logit} 
\newcommand{\T}{\mathrm{T}} 

\newcommand{\ESS}{\mathit{ESS}} 
\newcommand{\CESS}{\mathit{CESS}} 
\newcommand{\reals}{\mathbb{R}}%
\newcommand{\naturals}{\mathbb{N}}%
\newcommand{\EG}{e.g.\@ }
\newcommand{\IE}{i.e.\@ }

\newcommand{\WP}{w.p.\@ }

\newcommand{\male}{\mathrm{m}}
\newcommand{\female}{\mathrm{f}}
\newcommand{\adult}{\mathrm{A}}
\newcommand{\juvenile}{1}
\newcommand{\survivors}{{\mathit{sur}}}
\newcommand{\immigrants}{{\mathit{imm}}}

%% file: abbreviations.tex
\newacronym{ABC}{ABC}{approximate Bayesian computation}%
\newacronym{PMCMC}{PMCMC}{particle Markov chain Monte Carlo}%
\newacronym[\glslongpluralkey={(general state-space) hidden Markov models}]{HMM}{HMM}{(general state-space) hidden Markov model}%
\newacronym{ESS}{ESS}{effective sample size}%
\newacronym{CESS}{CESS}{conditional effective sample size}%
\newacronym{PDF}{PDF}{probability density function}%
\newacronym{IID}{IID}{independent and identically distributed}%
\newacronym{MCMC}{MCMC}{Markov chain Monte Carlo}%
\newacronym{MH}{MH}{Metropolis--Hastings}%
\newacronym{PMMH}{PMMH}{particle marginal Metro\-po\-lis--Has\-tings}%
\newacronym{CDF}{CDF}{cumulative distribution function}%
\newacronym{SMC}{SMC}{sequential Monte Carlo}%
\newacronym{CSMC}{CSMC}{conditional sequential Monte Carlo}%
\newacronym{RJMCMC}{RJMCMC}{reversible-jump Markov chain Monte Carlo}%
\newacronym{ML}{ML}{maximum likelihood}%
\newacronym{MLE}{MLE}{maximum-likelihood estimate}%
\newacronym{PF}{PF}{particle filter}%
\newacronym{BUGS}{BUGS}{Bayesian inference Using Gibbs Sampling}%
\newacronym{JAGS}{JAGS}{Just Another Gibbs Sampler}%
\newacronym{DA}{DA}{delayed-acceptance}%
\newacronym{MSE}{MSE}{mean-square error}%
\newacronym{WP}{w.p.}{with probability}%